\documentclass[%
 aip,
 amsmath,amssymb,
 reprint,%
]{revtex4-1}
\usepackage{graphicx}% Include figure files
\usepackage{dcolumn}% Align table columns on decimal point
\usepackage{bm}% bold math
\usepackage[utf8]{inputenc}
\usepackage[T1]{fontenc} % Use modern font encodings
\usepackage{mathptmx}
\usepackage{etoolbox}

\usepackage[version=3]{mhchem} % Formula subscripts using \ce{}    
\usepackage{comment}

\RequirePackage{epstopdf}
\DeclareGraphicsExtensions{.eps}
\usepackage[separate-uncertainty = true,multi-part-units=single]{siunitx}
\DeclareSIUnit{\wtpercent}{wt.\%}
\DeclareSIUnit{\Torr}{Torr}
\usepackage{mhchem}
\usepackage{booktabs}
\usepackage{svg}
\usepackage{natbib}
\usepackage{physics}

\usepackage{hyperref}
\usepackage{cleveref}
\crefname{figure}{Figure~}{Figures~}
\crefname{table}{Table~}{Tables~}

\usepackage{placeins}

\makeatletter
\def\@email#1#2{%
 \endgroup
 \patchcmd{\titleblock@produce}
  {\frontmatter@RRAPformat}
  {\frontmatter@RRAPformat{\produce@RRAP{*#1\href{mailto:#2}{#2}}}\frontmatter@RRAPformat}
  {}{}
}%
\makeatother
\begin{document}

\preprint{AIP/123-QED}

\title[]{Preparation and evaluation of alexandrite, forsterite, and topaz substrates for the epitaxial growth of rutile oxides}

% Force line breaks with \\
\author{Monique Kubovsky}
\homepage{contributed equally}
\affiliation{Platform for the Accelerated Realization, Analysis, and Discovery of Interface Materials (PARADIM), Cornell University, Ithaca, New York 14853, USA}
 
\author{Yorick A. Birkhölzer}
\homepage{contributed equally}
\affiliation{Department of Materials Science and Engineering, Cornell University, Ithaca, New York 14853, USA}
\email{y.birkholzer@cornell.edu}
\author{Luka B. Mitrovic}
\affiliation{Department of Materials Science and Engineering, Cornell University, Ithaca, New York 14853, USA}

\author{Hanjong Paik}
\homepage{Present address: School of Electrical and Computer Engineering, The University of Oklahoma, Norman, Oklahoma, USA}
\affiliation{Platform for the Accelerated Realization, Analysis, and Discovery of Interface Materials (PARADIM), Cornell University, Ithaca, New York 14853, USA}
\affiliation{Department of Materials Science and Engineering, Cornell University, Ithaca, New York 14853, USA}

\author{George R. Rossman}
\affiliation{Division of Geology and Planetary Science, California Institute of Technology, Pasadena, Califronia 91125, USA}

\author{Darrell G. Schlom}
\homepage{https://schlom.mse.cornell.edu.}
\affiliation{Platform for the Accelerated Realization, Analysis, and Discovery of Interface Materials (PARADIM), Cornell University, Ithaca, New York 14853, USA}
\affiliation{Department of Materials Science and Engineering, Cornell University, Ithaca, New York 14853, USA}
\affiliation{Kavli Institute at Cornell for Nanoscale Science, Ithaca, New York 14853, USA}
\affiliation{Leibniz-Institut für Kristallzüchtung, Max-Born-Str. 2, 12489 Berlin, Germany}

\date{\today}% It is always \today, today,

\begin{abstract}
%Metal-insulator transitions and superconductivity are fascinating phenomena that have great potential to revolutionize electronics. 
%However, the synthesis of thin films of materials that display these intriguing properties is severely limited by the lack of suitable substrates. 
%A prime example of such materials are transition metal oxides with the rutile structure.
%

Metal-insulator transitions and superconductivity in rutile-structured oxides hold promise for advanced electronic applications, yet their thin film synthesis is severely hindered by limited substrate options. Here, we present three single-crystalline substrates, \ce{BeAl2O4}, \ce{Mg2SiO4}, and \ce{Al2SiO4(\text{F,OH})_2}, prepared via optimized thermal and chemical treatments to achieve atomically smooth surfaces suitable for epitaxial growth. Atomic force microscopy confirms atomic step-and-terrace surface morphologies, and oxide molecular-beam epitaxy growth on these substrates demonstrates successful heteroepitaxy of rutile \ce{TiO2}, \ce{VO2}, \ce{NbO2}, and \ce{RuO2} films. Among these unconventional substrates, \ce{BeAl2O4} exhibits exceptional thermal and chemical stability, making it a versatile substrate candidate. These findings introduce new substrate platforms that facilitate strain engineering and exploration of rutile oxide thin films, potentially advancing the study of their strain-dependent physical properties.

% Commonly available rutile substrates are not stable under the conditions necessary for growing oxide thin films with a rutile crystal structure. 
%
%In this work, we present a set of  substrates that can be used for the hetero-epitaxial thin film growth of rutile thin films.
%We demonstrate how we optimized processes that prepare these substrates for thin films deposition.
%Atomic force microscopy shows regular and smooth step-and-terrace morphologies.
%
%We present case studies evaluating the suitability of these new substrates for epitaxial growth of rutile films via oxide molecular beam epitaxy.

\end{abstract}

% Keywords: Epitaxial growth, rutile structure, transition metal oxides, heteroepitaxy, atomic force microscopy

\maketitle

\section{\label{sec:level1}Introduction}

Experimental evidence for strain-induced superconductivity was found in epitaxial thin films of \ce{RuO2} under large compressive strain.\cite{Ruf2021, Wadehra2025}
% https://www.nature.com/articles/s43246-025-00856-6
%
This is a very remarkable result, since it was the first report of the induction of superconductivity in any material that does not superconduct in its bulk form, but does become a superconductor in thin film form by the application of epitaxial strain. 
This motivates us to expand the toolbox for strain engineering of more oxides with the rutile structure.
Conflicting reports regarding a possible altermagnetic ground state in \ce{RuO2} have further fueled interest in the synthesis of rutile oxides.\cite{Smejkal2022, Feng2022, Kessler2024, Liu2024, Occhialini2025, Gregory2025} 
In addition to the possibility of discovering superconductivity in rutile oxide thin films, rutiles are a target of research due to their metal-insulator transitions, such as those in \ce{VO2} and \ce{NbO2},\cite{Morin1959, Janninck1966, Belanger1974} as well as possible altermagnetism in \ce{ReO2}.\cite{Chakraborty2024} Transition metal rutile oxides are also interesting due to their catalytic properties,\cite{Reese2024} as well as the high spin-polarization of \ce{CrO2}.\cite{Soulen1998, Anguelouch2002} % MnO2, RuO2, IrO2 
A comprehensive overview of rutiles and birutiles has been published by \citet{Hiroi2015}.

Historically, much work in the oxide thin film community has focused on perovskites with the generalized chemical formula \ce{ABO3}.
Among the commercially available oxide single crystal substrates for thin film growth, the vast majority are perovskites or layered perovskites with the Ruddlesden-Popper structure with the generalized formula \ce{A2BO4}.\cite{Fratello2024}
Rutiles as well as the B-site of \ce{ABO3} perovskites consist of exclusively octahedrally coordinated metal cations.
A key difference between the perovskite and rutile structures is that in the former all octahedra are corner-sharing, whereas the hallmark of the rutile structure is an edge-sharing chain of octahedra along the rutile $c$-axis. This gives rutiles properties with strong anisotropy. A well-known example is the strong birefringence in \ce{TiO2}.

%For the olivine and corundum materials, the effective lattice parameters are deduced from a near-coincident-site lattice model matching the rectangular surface unit cell of rutile (101) are shown.

The only two commercially available substrates with the rutile crystal structure are \ce{TiO2} and \ce{MgF2}. 
In a research environment where sample synthesis is not limited by the thermal budget of back-end-of-line-compatible processes, there is a trend to explore the higher synthesis temperatures to improve structural perfection. This regime often enables adsorption controlled growth, which typically unlocks the best thin film properties.\cite{Driscoll2020}
% [J.L. MacManus-Driscoll, M.P. Wells, C. Yun, J.-W. Lee, C.-B. Eom, and D.G. Schlom, “New approaches for achieving more perfect
%transition metal oxide thin films,” APL Mater. 8(4), 040904–13 (2020).]
This approach recently took a massive step forward through the availability of powerful \ce{CO2} substrate laser heaters that can directly heat oxide substrates.\cite{Braun2020, Hensling2024}
%, [F.V.E. Hensling, W. Braun, D.Y. Kim, L.N. Majer, S. Smink, B.D. Faeth, and J. Mannhart, “State of the art, trends, and opportunities for oxide epitaxy,” APL Mater. 12(4), 040902 (2024).]
Regarding high-temperature stability, \ce{TiO2} is not an ideal substrate, as it rapidly reduces at elevated temperatures and low oxygen partial pressures, which are needed to stabilize intermediate-valence oxides such as \ce{TaO2}.
Even more dramatically, we found that \ce{MgF2} already decomposes around \SI{550}{\celsius}.
An additional challenge to the growth of an oxide (e.g., \ce{TiO2}) on \ce{MgF2} is the propensity of the substrate to be oxidized to form a volatile fluoride gas (e.g., \ce{TiF4}) through the reaction

\begin{equation}
    2~\ce{MgF2} + \ce{Ti} + \ce{O2} \rightarrow 2~\ce{MgO} + \ce{TiF4}.
\end{equation}
At \SI{800}{\kelvin}, this reaction is favorable with $\Delta G^0$ of \SI{-257.8}{\kilo\joule\per\mole} of \ce{MgF2}. \cite{Barin1995}
To epitaxially strain rutile thin films, more substrates with a rutile-like structure are thus needed.

The literature on the bulk single crystal growth of rutile substrates is sparse.
Two different rutiles, one with small (\ce{GeO2}) and one with large (\ce{SnO2}) lattice parameters, stand out in particular.
In the quest for ultrawide-bandgap semiconductors, \citet{Chae2022} demonstrated flux-grown \ce{GeO2} as a rutile substrate, albeit with rather small sample sizes up to 4 x \SI{2}{\milli\meter\squared}.
Recently, \citet{Galazka2025} demonstrated top-seeded solution growth of rutile \ce{GeO2} with diameters up to 15 mm from which he was able to successfully prepare substrates with a dimension of 5 x \SI{5}{\milli\meter\squared}. While \ce{GeO2} promises to be a good substrate for \ce{MnO2} and \ce{CrO2} with similarly small lattice parameters, here we are looking for new substrates with a larger lattice parameter.
% https://onlinelibrary.wiley.com/doi/full/10.1002/pssb.202400326
On that note, it is worth mentioning that  Galazka \textit{et al.} published the growth of rutile \ce{SnO2} by physical vapor transport (PVT) with diameters up to 25 mm and cut substrates with a dimension of 5 x \SI{5}{\milli\meter\squared}.\cite{Galazka2014}
% https://onlinelibrary.wiley.com/doi/full/10.1002/pssa.201330020
Despite the promising lattice mismatch of \ce{SnO2} to larger rutiles, it is not stable at high temperatures.\cite{Adkison2020} This makes it not suitable for the growth of materials such as \ce{TaO2}, which require temperatures around \SI{1000}{\celsius} for crystallization.

\begin{figure}[t]
    \centering
    \includegraphics[width=\columnwidth]{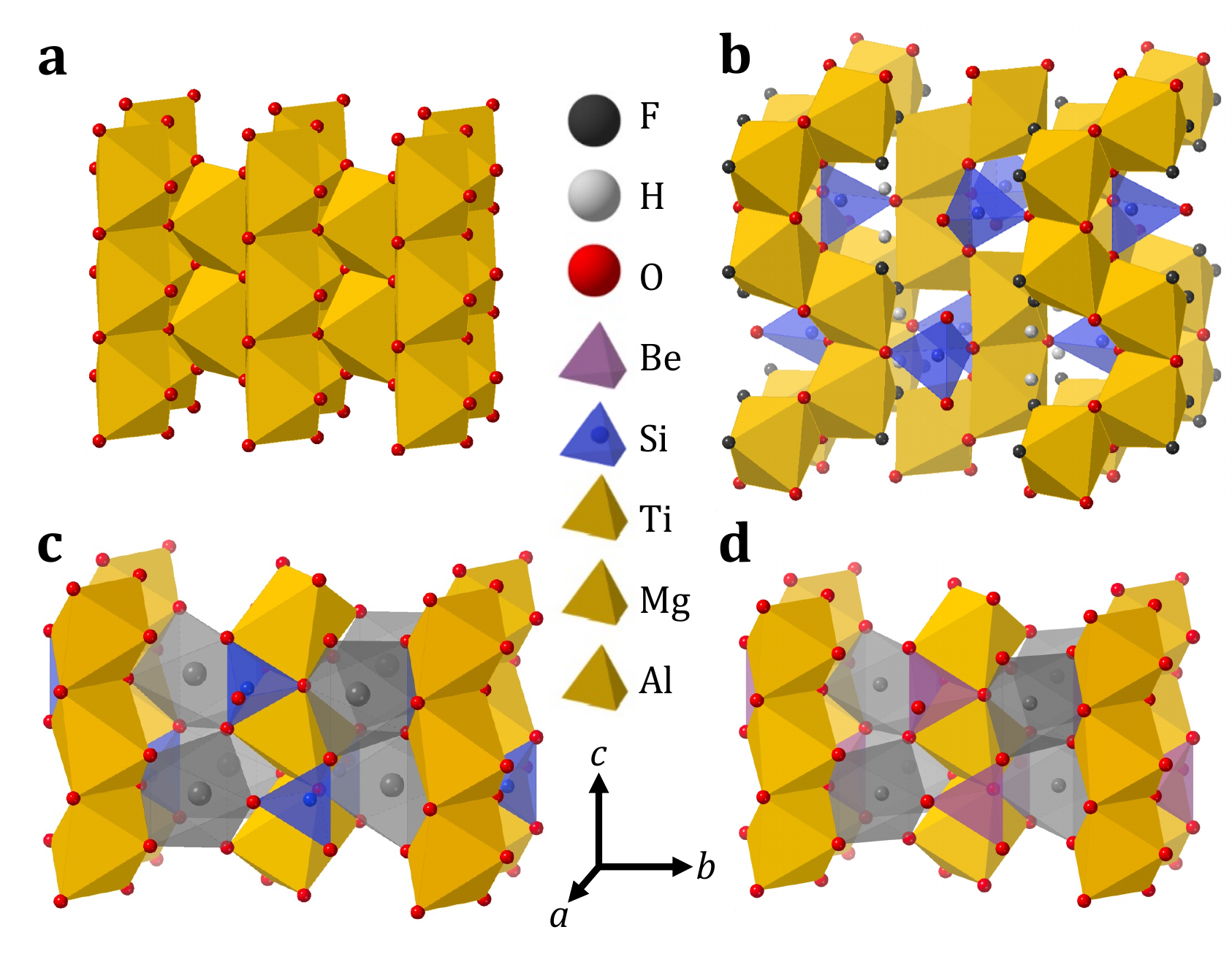}
    \caption{Schematics of the (a) rutile, (b) topaz, (c) forsterite, and (d) alexandrite crystal structures. Edge-sharing oxygen octahedra forming the common motif along the $c$-axis are highlighted in yellow. Octahedral Mg- and Al-sites outside the chains are shown in gray in (c) and (d).}
    \label{fig:overview}
\end{figure}

% At room temperature, you would expect the BCT rather than the undistorted rutile phase for NbO2

The ideal substrate for epitaxy is stable under both oxidizing and reducing conditions at elevated temperatures. This criterion motivated our search beyond the ubiquitous transition-metal oxides with multiple valence states.
While isostructural substrates would be ideal for the synthesis of rutile thin films, non-rutile substrates can, in principle, be used for heteroepitaxy. 
An early such example is the epitaxial synthesis of \ce{TiO2} on \ce{Al2O3} by \citet{Fukushima1993}. They used the concept of the near-coincidence-site lattice (NCLS) \cite{Balluffi1982}
% [R.W. Balluffi, A. Brokman, and A.H. King, “CSL/DSC lattice model for general crystalcrystal boundaries and their line defects,” Acta Metallurgica 30(8), 1453–1470 (1982).]
to explain the orientational relationships between the film and \ce{Al2O3} substrate.

While our group and others have confirmed that single-crystalline heteroepitaxy of (101)-oriented rutile-structured oxide thin films is possible on \ce{Al2O3} $(1\bar{1}02)$ ($r$-plane sapphire), the resulting layers are highly anisotropically strained at best. While one in-plane direction matches commensurately (as for example in the case of \ce{TaO2} on \ce{Al2O3} $(1\bar{1}02)$), %\cite{Birkholzer2025}, % if TaO2 paper comes out first
the orthogonal in-plane direction suffers from a very large ($\approx$ \SI{10}{\%}) mismatch.

To expand the playing field for rutile epitaxy, in this article we present three materials that can serve as substrates for rutile heteroepitaxy: \ce{BeAl2O4}, \ce{Mg2SiO4}, and \ce{Al2SiO4(\text{F,OH})_2}.
In \cref{tab:lattice_parameters}, the lattice parameters of these orthorhombic materials are summarized.
In Fig. \ref{fig:overview}, the rutile structure is depicted alongside topaz, forsterite, and alexandrite to highlight the similar structural motif shared by all structures: chains of edge-sharing oxygen octahedra along the $c$-axes.

% Figure 0

%For comparison, the table also lists the lattice parameters of two rutile oxide thin films, \ce{RuO2} and \ce{NbO2}\footnote{For simplicity, in this article we refer to \ce{NbO2} in its high-temperature rutile structure. At room temperature, bulk \ce{NbO2} forms a superstructure that has a four times larger unit cell}.

\begin{table}[t]
\setlength{\tabcolsep}{6pt}
    \begin{center}
    \caption{Lattice parameters of rutile, forsterite, alexandrite, and topaz.}
    \begin{tabular}{llllll}
       \toprule
        Compound & $a$ (\SI{}{\angstrom}) & $b$ (\SI{}{\angstrom}) & $c$ (\SI{}{\angstrom}) &   Reference \\
       \midrule
        \ce{TiO2}                & 4.594 & 4.594  & 2.959 &  \cite{Baur1971} \\
        \ce{Mg2SiO4}             & 4.753 & 10.199 & 5.981 &  \cite{Lager1981} \\
        \ce{BeAl2O4}             & 4.429 & 9.407  & 5.478 &  \cite{Dudka1985} \\
        \ce{Al2SiO4(\text{F,OH})_2} & 4.667 & 8.834  & 8.395 &  \cite{Gatta2006} \\
       \bottomrule
    \end{tabular}
    \label{tab:lattice_parameters}
\end{center}
\end{table}

\begin{table}[b]
\setlength{\tabcolsep}{6pt}
    \begin{center}
    \caption{Expected out-of-plane orientation of film and substrate pairs.}
    \begin{tabular}{ll}
        \toprule
        Substrate & Film \\
        \midrule
        Alexandrite or forsterite (010) & Rutile (101) or (010) \\
        Alexandrite or forsterite (001) & Rutile (001) \\
        Topaz (010) & Rutile (010) \\
        Topaz (001) & Rutile (001) \\
        \bottomrule
    \end{tabular}
    \label{tab:orientation_matching}
    \end{center}
\end{table}

A prerequisite for optimal epitaxial thin film growth is a high-quality crystalline substrate with an atomically flat surface with controlled surface termination.\cite{Biswas2017}
We therefore present thermal and chemical treatments to prepare such surfaces.

\ce{BeAl2O4} is known under the mineral names chrysoberyl and alexandrite (when Cr-doped). We purchased the substrates from Northrop Grumman Synoptics, a US-based manufacturer of laser crystals. The \SI{0.08}{\wtpercent} Cr-doped \ce{BeAl2O4} was grown by the Czochralski method.
The Czochralski growth of \ce{BeAl2O4} dates back to the 1960s,\cite{Cline1979} and commercial alexandrite lasers ($\lambda$ = \SI{755}{\nano\meter}) are well-established for applications such as hair and tattoo removal  since the 1990s.\cite{Finkel1997, Fitzpatrick1994} In contrast to the next two materials we introduce in this article, there are no reports of using \ce{BeAl2O4} as a substrate for epitaxial growth. 
The high structural quality of these crystals is reflected in a narrow rocking curve full width at half maximum (FWHM) of 15 arcsec. %data not shown

\ce{Mg2SiO4} is known under its mineral name forsterite. We purchased the Czochralski-grown raw material from Oxide Corp, Japan, and had it diced and polished by the German company SurfaceNet. The Czochralski growth of \ce{Mg2SiO4} was described in the early 1970s.\cite{Finch1971, Takei1974}
In this work, we refer to forsterite and isostructural alexandrite using the non-standard (but most commonly used) \textit{Pbnm} setting of space group \#62, which leads to matching Miller indices for several orientations of rutile adlayers as well as the substrate.  
\footnote{An equivalent notation can be obtained through permutation in the space group \textit{Pnma}.}

The topaz used in this study was of natural origin and oriented and polished by the German vendor CrysTec.
The reliance on a natural mineral stems from the impossibility of growing topaz from the melt and the insufficient size of early hydrothermally grown crystals,\cite{Somiya1989} which were at least two orders of magnitude too small for substrate applications.

%\citet{Somiya1989} used the hydrothermal method to grow very small crystals of topaz whose size remains at least 2 orders of magnitude too small to be used as a substrate for thin film growth. 
%More recently, \citet{Trujillo-Vazquez2017} used chemical vapor deposition to grow topaz crystals, with sizes that remained on the micrometer scale.
%https://doi.org/10.1016/j.jallcom.2017.01.159

A significant recent advance was reported by \citet{Setkova2024}, who not only synthesized mm-scale synthetic topaz but also achieved a controlled expansion of the lattice parameter by up to \SI{3}{\percent} through the partial substitution of Al with Ga and Si with Ge. Crucially, however, these crystals were grown on natural topaz seeds, and their resulting size and quality were insufficient to meet the demands posed on substrates for epitaxy.

All substrates used in this study were available in dimensions up to 10x10x\SI{1}{\milli\meter\cubed}.
The expected orientations of the films for various substrate orientations are summarized in \cref{tab:orientation_matching} and will be discussed in detail in the next section.

\begin{figure}[t]
    \centering
    \includegraphics[width=\columnwidth]{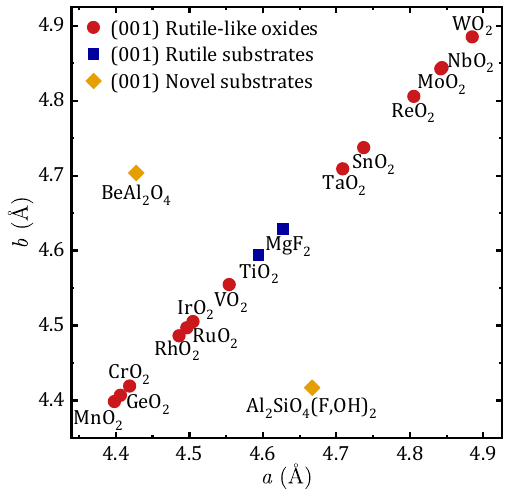}
    \caption{In-plane lattice parameters of select (001)-oriented rutile-like oxides, commercially available rutile substrates, alexandrite, and topaz.}
    \label{fig:lattice_parameters_001}
\end{figure}

% NbO2 lattice parameters are simplified from BCT structure at room temperature

%\FloatBarrier
\section{\label{sec:cartoons}Orientational match of alexandrite, forsterite, and topaz substrates with rutile}

\begin{figure}[t]
    \centering
    \includegraphics[width=\columnwidth]{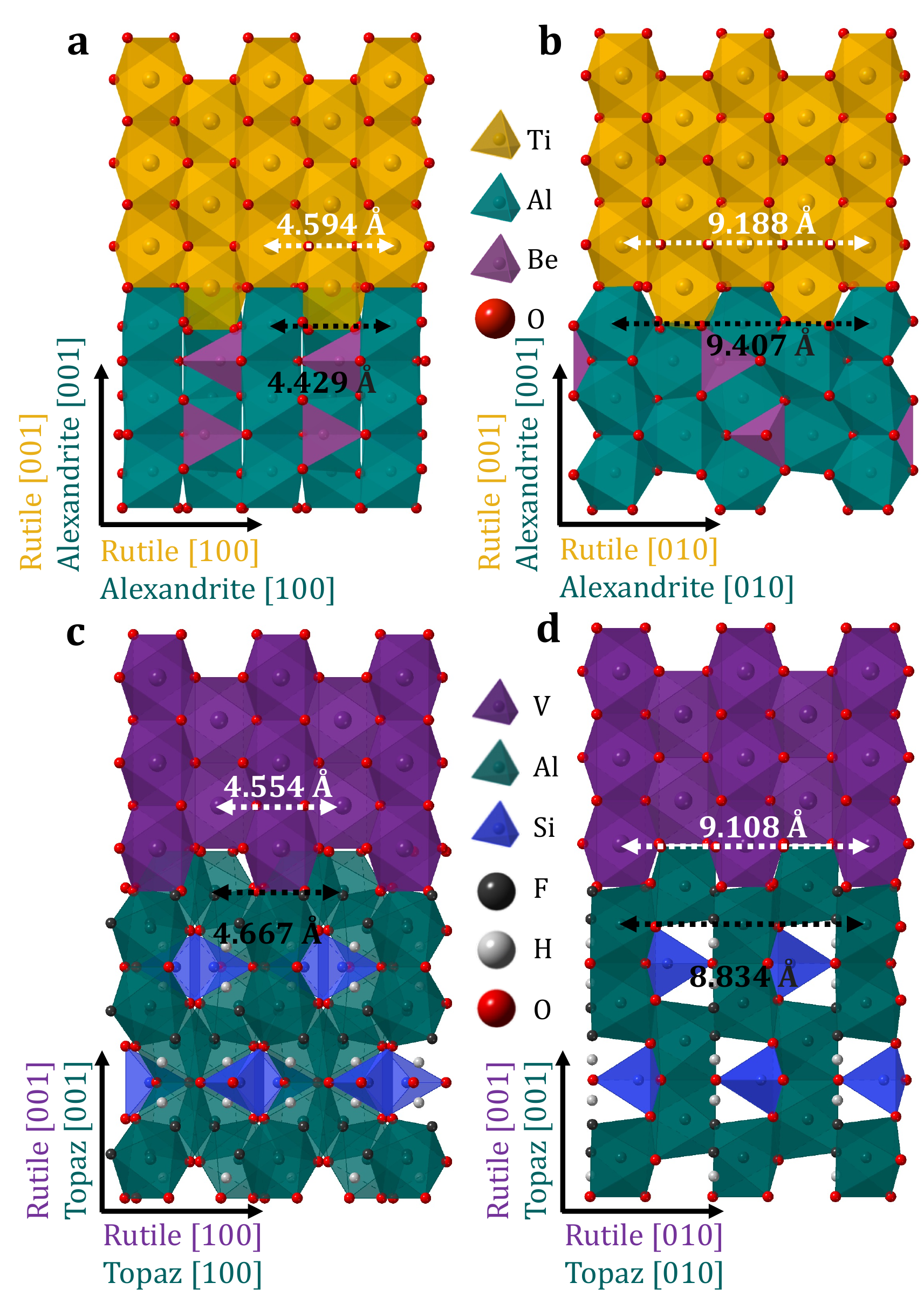}
    \caption{Cross-sectional schematic of \ce{TiO2} (001) on \ce{BeAl2O4} (001) viewed along the substrate [100] (a) and [010] directions (b); \ce{VO2} (001) on \ce{Al2SiO4(\text{F,OH})_2} (001) viewed along the substrate [001] (c) and [010] direction (d).}
    \label{fig:RuO2_001_cartoon}
\end{figure}

In this article, we present multiple orientations of three different substrates for rutile epitaxy, leading to multiple possible epitaxial interfaces. Our choice of substrates is guided by three factors: (1) the availability of single crystals with excellent structural quality and of sufficient size to yield 10x\SI{10}{\milli\meter\squared} substrates, (2) a similar structural motif to a rutile so that what we call "local epitaxy" can occur during growth, and (3) an NCSL with a high concentration of NCSL sites at the interface between substrate and film and small lattice mismatch. By local epitaxy we refer to sub-unit cell structural elements that are similar in the film and substrate and allow depositing atoms to continue the structural motif between substrate and film via epitaxy. We particularly focus on continuity in the type and arrangement of coordination polyhedra. This can be viewed as following Pauling's rules \cite{Pauling1929}
% [L. Pauling, “The principles determining the structure of complex ionic crystals,” J. Am. Chem. Soc. 51, 1010–1026 (1929).] 
across interfaces. \cite{Andeen2007}
% [D. Andeen and F.F. Lange, "Crystal chemistry of interfaces formed between two different non-metallic, inorganic structures," Int. J. Mat. Res. 98, 1222–1229 (2007)] 
The specific substrates and orientations were deliberately selected for the growth of rutile films by carefully considering their crystal structures and as we report all were found to provide epitaxial growth for rutile films. A striking feature of the rutile structure is the chains of edge sharing \ce{TiO6} octahedra it contains. It is this structural motif that was the focus of our local epitaxy approach. In selecting potential substrates, we selected oxide (or oxyfluoride) substrates with this same structural element. The similarity in structures is evident in Fig. \ref{fig:overview}, where the edge-sharing chains of octahedra are colored yellow in rutile and the three substrates selected for this study.
Inspired by the NCLS model,\cite{Balluffi1982}
% [R.W. Balluffi, A. Brokman, and A.H. King, “CSL/DSC lattice model for general crystalcrystal boundaries and their line defects,” Acta Metallurgica 30(8), 1453–1470 (1982).]
we designed heterostructures wherein the oxygen octahedral network across the interface between substrate and film is preserved as best as possible. In the following subsections, we compare the effective in-plane lattice parameters of bulk-like, unstrained film and substrate pairings and schematically sketch the expected cross-sectional interfaces. \footnote{In the case of pseudocubic perovskites, a single number line suffices to compare lattice parameters of different materials. In the case of tetragonal rutile thin films and highly orthorhombic substrates, one representation does not suffice and several orientations need to be examined individually.}

\begin{figure}[t]
    \centering
    \includegraphics[width=\columnwidth]{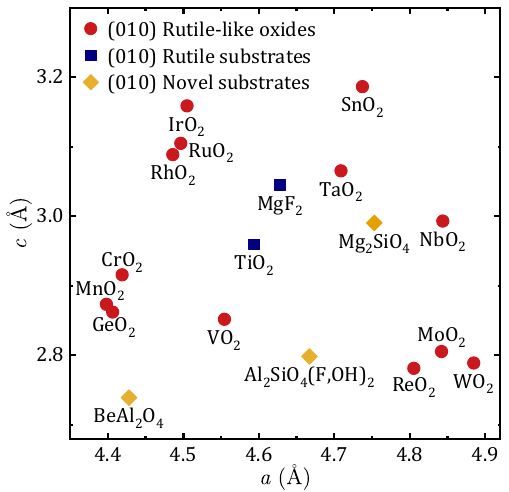}
    \caption{In-plane lattice parameters of rutile (010) materials compared with commercially available rutile substrates, alexandrite, forsterite, and topaz.}
    \label{fig:lattice_parameters_010}
\end{figure}

\subsection{Substrates for (001) rutile thin films}
Let us start with (001)-oriented films, where for an undistorted rutile crystal structure, the in-plane directions are the indistinguishable $a$ and $b$-axes, respectively. Fig. \ref{fig:lattice_parameters_001} compares a selection of rutile oxide lattice parameters \cite{Hiroi2015} with \ce{BeAl2O4} (001) and topaz (001). For simplicity, we ignore the fact that some rutile-like oxides such as \ce{MoO2} and \ce{NbO2} have polymorphs with slightly distorted crystal structures with lower symmetry than the idealized rutile. In this article, we discuss pseudotetragonal $a$ = $b$ for all rutile-like oxides.

In Fig. \ref{fig:RuO2_001_cartoon}, we show the interface between rutile \ce{TiO2} (001) and \ce{BeAl2O4} (001) as well as between rutile \ce{VO2} (001) and topaz (001) in a cross-sectional schematic.
Additional perspectives along the <110> directions are illustrated in Fig. \ref{fig:SI_001_along110}.
The geometric match between the substrate and thin film at the interface suggests that high quality epitaxial rutile thin films can be grown on \ce{BeAl2O4} (001). An example is shown later in this article.
Being able to template (001) rutile growth is particularly exciting since this plane has a very high surface energy compared to the (010) and (101) orientations, which are discussed in the following subsections. \footnote{The (010) orientation is indistinguishable from the (100) orientation in undistorted rutiles, but we chose this notation for a convenient comparison with the orthorhombic substrates.}

\begin{figure}[t]
    \centering
    \includegraphics[width=\columnwidth]{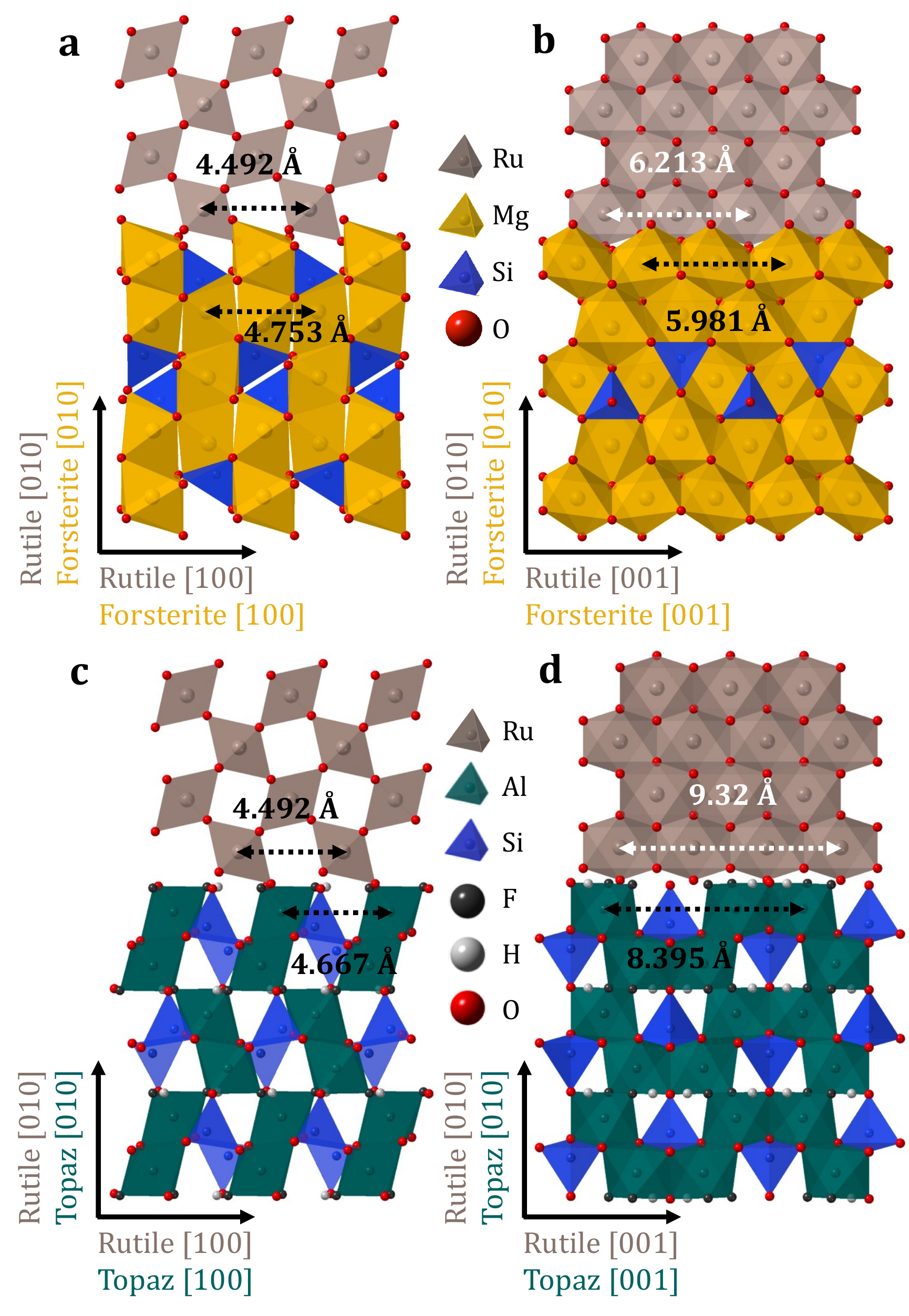}
    \caption{Cross-sectional schematic of  \ce{RuO2} (010) on \ce{Mg2SiO4} (010) (a,b) and \ce{RuO2} (001) on \ce{Al2SiO4(\text{F,OH})_2} (001) viewed along the substrate [001] (a) and [100] directions (c,d).}
    \label{fig:RuO2_Topaz_010_cartoon}
\end{figure}

\subsection{Substrates for (010) rutile thin films}
In the case of (010)-oriented rutile growth, the in-plane lattice parameters are the $a$ and $c$ axes.
Fig. \ref{fig:lattice_parameters_010} compares select rutile lattice parameters\cite{Hiroi2015} with \ce{BeAl2O4} (010), \ce{Mg2SiO4} (010), and topaz (010).
In Fig. \ref{fig:RuO2_Topaz_010_cartoon}, we schematically show the interface between \ce{RuO2} (010) and \ce{Mg2SiO4} (010), as well as \ce{RuO2} (010) and topaz (010).

Preliminary attempts to grow \ce{RuO2} on \ce{Mg2SiO4} (010) led to (010)-oriented rutile growth despite the large lattice mismatch. 
This result is particularly interesting because the rutile $c$-axis lies in-plane in this geometry and can potentially be subject to a large compressive strain.

\subsection{Substrates for (101) rutile thin films}
\begin{figure}[t]
    \centering
    \includegraphics[width=\columnwidth]{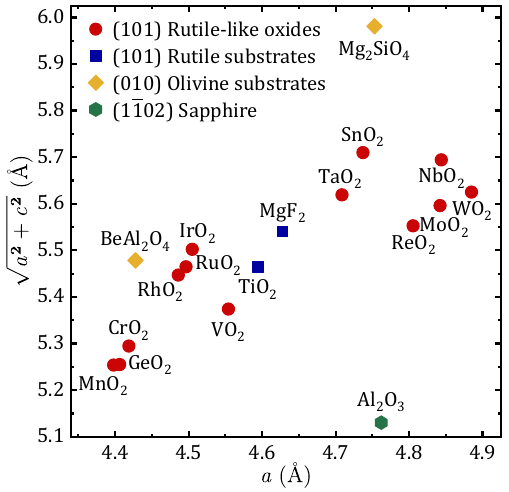}
    \caption{In-plane lattice parameters of select (101)-oriented rutile-like oxides, commercially available rutile and sapphire substrates, alexandrite, and forsterite. Note that for undistorted rutile, $a$ = $b$.}
    \label{fig:lattice_parameters_101}
\end{figure}

%% 101
Finally, we examine the case of (101)-oriented rutile growth.
For such films, the in-plane directions are $[10\bar{1}]$ and $[010]$ with effective lengths $\sqrt{a^2 + c^2}$ and $b$, respectively. 
Fig. \ref{fig:lattice_parameters_101} summarizes the resulting in-plane parameters of select rutile oxides \cite{Hiroi2015} and compares them with \ce{BeAl2O4} and \ce{Mg2SiO4} (010) as well as \ce{Al2O3} $(1\bar{1}02)$.

\begin{figure}[t]
    \centering
    \includegraphics[width=\columnwidth]{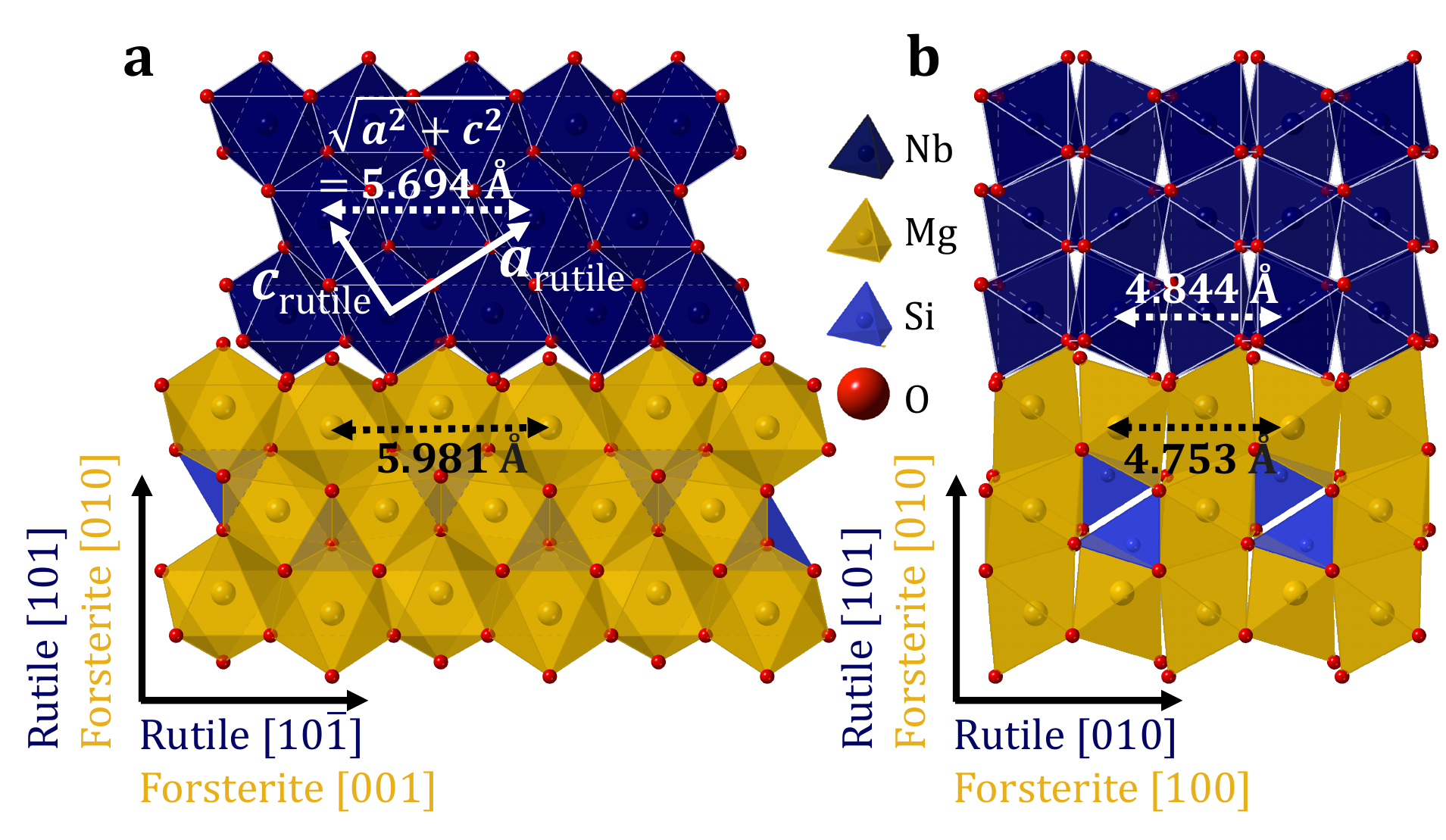}
    \caption{Cross-sectional schematic of rutile \ce{NbO2} (101) on \ce{Mg2SiO4} (010) projected along [100] (a) and [001] (b).}
    \label{fig:NbO2_Mg2SiO4_cartoon}
\end{figure}

Fig. \ref{fig:NbO2_Mg2SiO4_cartoon} shows the expected interface between \ce{Mg2SiO4} (010) and \ce{NbO2} (101), which we experimentally verified, see Section \ref{subsec:xrd}.
The same orientational relationship was also experimentally verified for \ce{RuO2} (101) on \ce{BeAl2O4} (010) (not shown).
We note that the forstertie (010) substrates can thus template both rutile (010) as well as rutile (101) epitaxial growth depending on the lattice mismatch.
% this is not ideal, but at least the samples were single-oriented in both case studies

Having outlined the expected orientational relationships based on bulk lattice parameters and in schematic form, in the next section, we present experimental annealing results of the aforementioned substrates, showing that they can be prepared with an atomically smooth step-and-terrace surface morphology as is desired for epitaxial overgrowth.
The development of substrate termination recipes that provide smooth step-and-terrace surfaces with known termination has been a major boon to the growth of oxide films on perovskite, \cite{Braun2020, Kawasaki1994, Koster1998, Ohnishi1999, Chang2008, Ngai2010, Kleibeuker2010, Biswas2011, Blok2011, Kleibeuker2012, Biswas2017, TOmar2018}
rocksalt,\cite{Braun2020}
rutile, \cite{Yamamoto2005} 
%[Y. Yamamoto, K. Nakajima, T. Ohsawa, Y. Matsumoto, and H. Koinuma, “Preparation of Atomically Smooth TiO2 Single Crystal Surfaces and Their Photochemical Property,” Jpn. J. Appl. Phys. 44(No. 17), L511–L514 (2005).]
sapphire, \cite{Yoshimoto1995, Smink2024, Brucker2025}
% [M. Yoshimoto, T. Maeda, T. Ohnishi, H. Koinuma, O. Ishiyama, M. Shinohara, M. Kubo, R. Miura, and A. Miyamoto, “Atomic-scale formation of ultrasmooth surfaces on sapphire substrates for high-quality thin-film fabrication,” Appl. Phys. Lett. 67(18), 2615–2617 (1995). 
% S. Smink, L.N. Majer, H. Boschker, J. Mannhart, and W. Braun, “Long-Range Atomic Order on DoubleStepped Al2O3(0001) Surfaces,” Adv. Mater. 36(24), e2312899 (2024). 
% M. Brucker, V. Harbola, J. Mannhart, S. Smink, T.J. Whittles, and F.V.E. Hensling, “Morphology of various single faced sapphire surfaces prepared by rapid thermal annealing,” Appl. Surf. Sci. 696, 162929 (2025).]
and cubic zirconia \cite{Ohta1999, Nakamura2007}
% [H. Ohta, H. Tanji, M. Orita, H. Hosono, and H. Kawazoe, “Heteroepitaxial Growth of Zinc Oxide Single Crystal Thin Films on (111) Plane YSZ by Pulsed Laser Deposition,” MRS Proc. 570(1), 309 (1999). 
% T. Nakamura, Y. Tokumoto, R. Katayama, T. Yamamoto, and K. Onabe, “RF–MBE growth and structural characterization of cubic InN films on yttria-stabilized zirconia (001) substrates,” J. Cryst. Growth 301, 508–512 (2007).
substrates.
Our work extends this ability to rutile-related substrates.

We use atomic force microscopy to image the surface of the substrates before and after thermo-chemical treatment.
Our preferred treatment is an \textit{in situ} thermal annealing step inside the growth chamber, which was performed at a background pressure of distilled ozone of \SI{1E-6}{\Torr} to prevent unintentional substrate reduction at elevated temperatures.
See the Methods section for further details.

\FloatBarrier
\section{\label{sec:results}Results}

\subsection{\label{sec:alexandrite}\ce{BeAl2O4}}

\begin{figure}[b]
    \centering
    \includegraphics[width=\columnwidth]{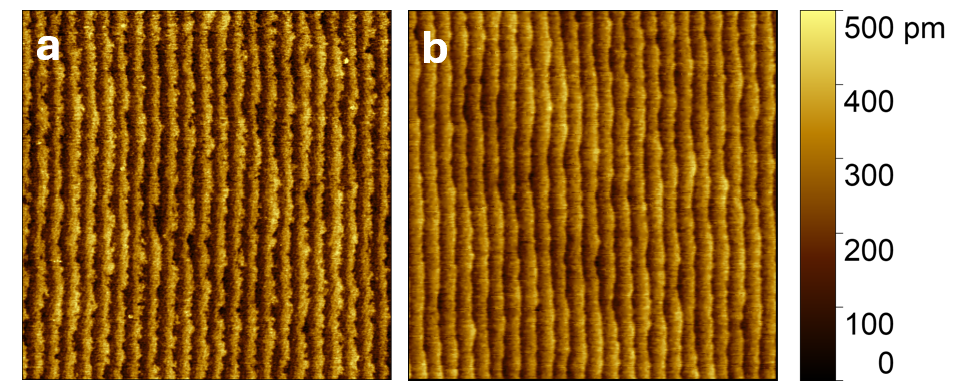}
    \caption{2 x \SI{2}{\micro\meter} AFM scans of (001) \ce{BeAl2O4} (a) in the as-received state and (b) after laser annealing at \SI{1250}{\celsius} for \SI{200}{\second}.}
    \label{fig:001BAO_B}
\end{figure}

Beginning with the (001) orientation of \ce{BeAl2O4}, the as-received substrate in Fig. \ref{fig:001BAO_B} (a) presents a step-and-terrace surface, where the terraces are of uniform width but have meandering step edges. Out of the subset of preparation procedures we used, the best results were produced by laser annealing at \SI{1250}{\celsius} for \SI{200}{\second}, shown in Fig. \ref{fig:001BAO_B} (b). Higher annealing temperatures (Fig. \ref{fig:001BAO_supp}) caused holes to form within the terraces, as well as excessive step meandering. Additionally, the procedure reduced the rms roughness from less than \SI{0.2}{\nano\meter} to less than \SI{0.1}{\nano\meter}. Annealing produced steps approximately \SI{0.25}{\nano\meter} high, which corresponds to 1/2 of the unit cell height in the [001] direction. Fig. \ref{fig:001BAO_steps} presents a possible model describing this behavior.

\begin{figure}[t]
    \centering
    \includegraphics[width=\columnwidth]{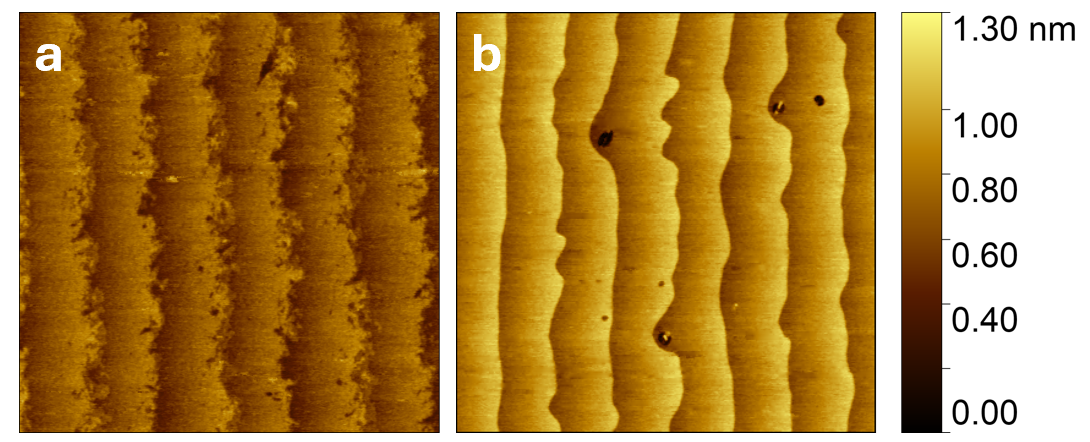}
    \caption{2 x \SI{2}{\micro\meter} AFM scans of (010) \ce{BeAl2O4} (a) in the as-received state and (b) after furnace annealing at \SI{900}{\celsius} for \SI{1}{\hour}.}
    \label{fig:010BAO_B}
\end{figure}

The as-received (010) \ce{BeAl2O4} surface is similarly rough to the (001) surface. Laser annealing for \SI{200}{\second} revealed that this plane becomes doubly-terminated, signified by triangular holes that change orientations on alternating terraces (Fig. \ref{fig:010BAO_supp}). As the desired step-and-terrace surface morphology was not achieved through rapid \textit{in situ} laser annealing, we used a tube furnace (Fig. \ref{fig:010BAO_T_supp}) to explore lower temperatures and longer annealing times\footnote{We switched to a tube furnace to avoid long annealing times inside the MBE growth chamber. There is no fundamental reason that prohibits long-duration \textit{in situ} annealing processes. A key difference, however, is the attainable gas pressure. While pressures higher than \SI{1E-5}{\Torr} are difficult to achieve in our vacuum tools, 1 atm oxygen pressure can readily be obtained in a tube furnace. The main advantage of \textit{ex situ} tube furnace annealing is the ability to prepare many sample simultaneously in parallel.}.
By furnace annealing at \SI{900}{\celsius} for an hour in 1 atm oxygen pressure, we achieved very smooth, uniform terraces and straightened steps with minimal meandering (Fig. \ref{fig:010BAO_B}). The rms roughness of both the as-received and annealed states were below \SI{0.2}{\nano\meter}. Annealing at higher temperatures causes raised needle structures with surrounding pits to form. 
Step height measurements reveal that the (010) \ce{BeAl2O4} surface becomes doubly-terminated through the formation of steps of fractional unit-cell height. Some samples had step heights between \SI{0.35}{\nano\meter} and \SI{0.4}{\nano\meter}, while others had heights around \SI{0.55}{\nano\meter}. These values roughly correspond to 2/5 and 3/5 of the unit cell height. While the exact surface termination is currently unknown, we present two possible models describing this behavior in Fig. \ref{fig:010BAO_steps}. To reduce inhomogeneity and the tendency of some films to form twins, a single-terminated surface would be more desirable. Future work could explore the influence of vicinal miscut angle and more finely tuned annealing conditions to achieve additional control of the step height.

%Due to the layered nature of the olivine structure in the [010] direction, the substrates prepared in this article exhibit a double termination, altering with each step.

\subsection{\label{sec:forsterite}\ce{Mg2SiO4}}
\begin{figure}[b]
    \centering
    \includegraphics[width=\columnwidth]{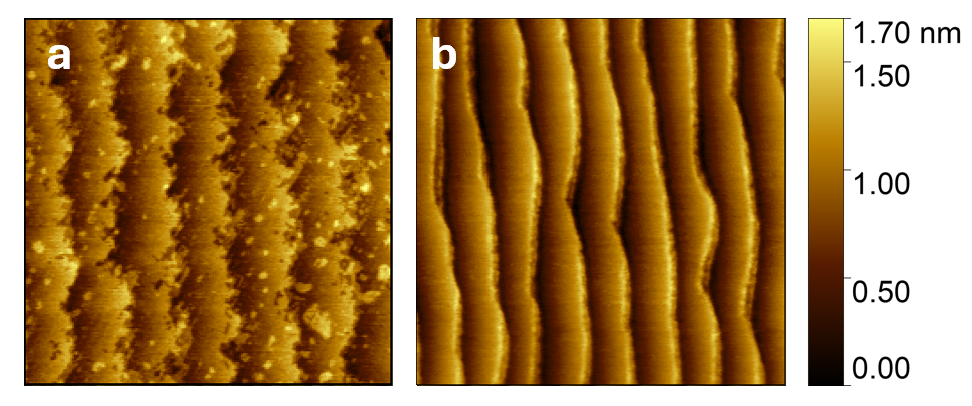}
    \caption{2 x \SI{2}{\micro\meter} AFM scans of (010) \ce{Mg2SiO4} (a) in the as-received state and (b) after laser annealing at \SI{1300}{\celsius} for \SI{200}{\second}.}
    \label{fig:010MSO_B}
\end{figure}

Fig. \ref{fig:010MSO_B} shows AFM scans before and after laser annealing at \SI{1300}{\celsius} for \SI{200}{\second}. This procedure adequately smoothed out the steps, maintained terrace width uniformity, and produced minimally waved, parallel steps. 
The rms roughness remained  $\approx $ \SI{0.3}{\nano\meter} as a result of the annealing. The prepared samples had step heights around \SI{1}{\nano\meter}, which is in very good agreement with the expected height of a single-unit-cell step. 
Results from additional annealing procedures at other temperatures are shown in Fig. \ref{fig:010MSO_supp}.
Due to limited availability of high-quality raw material, other orientations of \ce{Mg2SiO4} have not been explored thus far.

\subsection{\label{sec:topaz}\ce{Al2SiO4(\text{F,OH})_2}}
Unlike the laboratory-grown substrates presented in the previous sections, the topaz used in this study was of natural origin. While the artificial growth of topaz has been attempted via the hydrothermal method, the dimensions of the crystals reported in the literature thus far remain too small for thin film studies.\cite{Somiya1989} Another challenge is the lack of compositional control regarding the F/OH ratio in natural or lab-grown topaz. We expect that the high fugacity of F- and OH-containing species, e.g., HF and \ce{H2O},  pose a serious limitation on the accessible temperature range for thin film growth.
% this is bad since we typically want to grow films by MBE as hot as we can to access an adsorption-controlled growth regime

Here, we use substrates of 1 mm thickness and 3 x 5, 5 x 5, and 10 x \SI{10}{\milli\meter\squared} lateral dimensions.
Both orientations of topaz were etched in a buffered HF solution for \SI{60}{\second} prior to any attempted annealing. Even without thermal annealing, this step alone already drastically improves the surface quality of topaz \footnote{\ce{BeAl2O4} is insoluble in all acids, preventing us from exploring chemical treatment alternatives to the high-temperature annealing}.

\begin{figure}[t]
    \centering
    \includegraphics[width=\columnwidth]{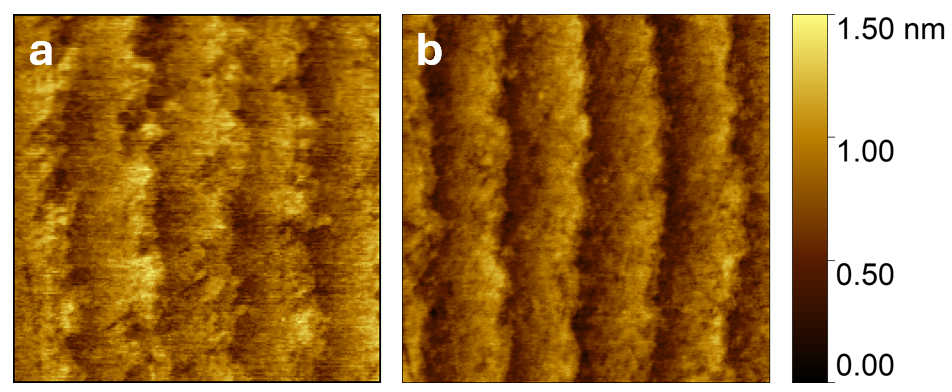}
    \caption{2 x \SI{2}{\micro\meter} AFM scans of (001) \ce{Al2SiO4(\text{F,OH})_2} (a) in the as-received state and (b) after furnace annealing at \SI{750}{\celsius} for \SI{1}{\hour}.}
    \label{fig:001Topaz_B}
\end{figure}

\begin{figure}[b]
    \centering
    \includegraphics[width=\columnwidth]{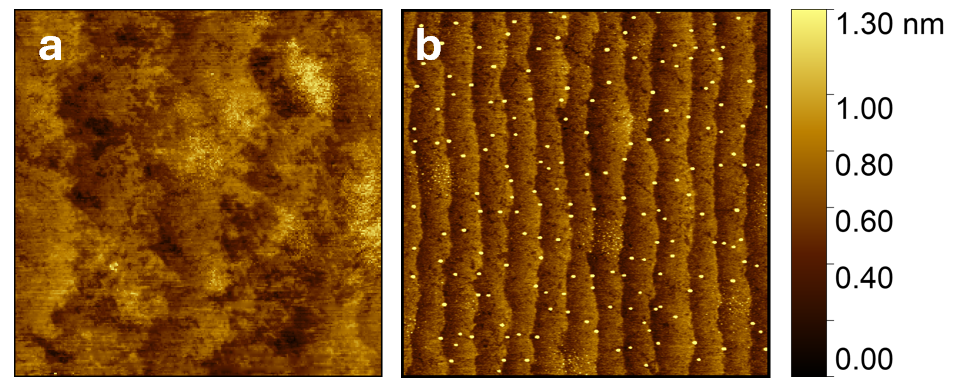}
    \caption{5 x \SI{5}{\micro\meter} AFM scans of (010) \ce{Al2SiO4(\text{F,OH})_2} (a) in the as-received state and (b) after furnace annealing at \SI{750}{\celsius} for \SI{1}{\hour}.}
    \label{fig:010Topaz}
\end{figure}
For the (001) surface, laser annealing at \SI{700}{\celsius} and \SI{800}{\celsius} for \SI{200}{\second} resulted in changes insurface morphology, but with little change in rms roughness compared with the as-received substrate (Fig. \ref{fig:001Topaz_supp}). At high temperatures the RHEED patterns began to fade, suggesting that the fluorine may be leaving the surface. Laser annealing at lower temperatures before the RHEED patterns deteriorate fails to substantially change the surface beyond its as-received state. To further investigate the lower anneal temperatures, we used a tube furnace. Furnance annealing at \SI{750}{\celsius} for an hour successfully produced much smoother steps than the as-received samples (Fig. \ref{fig:001Topaz_B}, \ref{fig:001Topaz_T_supp}). The rms roughness of both the as-received and annealed states was below \SI{0.2}{\nano\meter}. The annealed substrates had step heights around \SI{0.8}{\nano\meter}, which matches the unit cell height of (001) topaz.

Acid etching the (010) topaz defined the step-and-terrace structure and left the steps much smoother than the as-received state. Unfortunately, the laser annealed samples had rough, meandering steps (Fig. \ref{fig:010Topaz_supp}).
Similar issues with the RHEED patterns were observed for the (010) surface as for the (001) orientation.
Annealing in a tube furnace at 1 atm oxygen background at lower temperatures with longer hold times allowed the steps to straighten out but caused particles to form across the surface (Fig. \ref{fig:010Topaz}, \ref{fig:010Topaz_T_supp}). It is possible that these particles could be removed with repeated etching in HF acid. The rms roughness of the annealed state increased to <\SI{0.6}{\nano\meter} from <\SI{0.2}{\nano\meter}, likely due to the particles that formed across the surface. Additionally, this preparation produced steps of \SI{0.45}{\nano\meter}, which corresponds to 1/2 of the unit cell height in the [010] direction.
At higher annealing temperatures, the surface is terminated with \SI{0.22}{\nano\meter} high steps that each have the height of 1/4 of the unit cell. A model of how this quadruple termination may occur is shown in Fig. \ref{fig:010Topaz_steps}.

When trying to capture RHEED images of topaz substrates, we frequently observed image-distorting charging artifacts that could not be mitigated by changing the ozone background pressure or increasing the substrate temperature. We ascribe this to the highly electrically insulating nature of topaz.

Attempts to anneal at even higher temperatures >> \SI{800}{\celsius} led to the irreversible destruction of the surface and the observation of circular arcs in the RHEED pattern, indicative of polycrystallinity. 

% Hampar and Zussman book chapter
\citet{Hampar1984} comprehensively reviewed the thermal breakdown of topaz, suggesting that breakdown begins above \SI{850}{\celsius} in air. 
The authors also reported easy damage of topaz by focused electron beams.\cite{Hampar1983}
We observed RHEED patterns of topaz to fade above \SI{600}{\celsius}, which could be due to a combination of electron-beam induced damage as well as temperature.
Of the substrate materials introduced in this article, topaz has the lowest thermal and chemical stability. Fortunately, many rutile oxides can be grown at low temperatures < \SI{400}{\celsius}. \cite{Ruf2021, Wadehra2025, Tashman2014, Paik2015}
%  [1,2, J.W. Tashman, J.H. Lee, H. Paik, J.A. Moyer, R. Misra, J.A. Mundy, T. Spila, T.A. Merz, J. Schubert, D.A. Muller, P. Schiffer, and D.G. Schlom, “Epitaxial Growth of VO2 by Periodic Annealing,” Applied Physics Letters 104 (2014) 063104. 
% H. Paik, J.A. Moyer, T. Spila, J.W. Tashman, J.A. Mundy, E. Freeman, N. Shukla, J.M. Lapano, R. Engel-Herbert, W. Zander, J. Schubert, D.A. Muller, S. Datta, P. Schiffer, and D.G. Schlom, “Transport properties of ultra-thin VO2 films on (001) TiO2 grown by reactive molecular-beam epitaxy,” Appl. Phys. Lett. 107(16), 163101 (2015).]

\subsection{\label{subsec:xrd}Thin film growth}
Having demonstrated successful annealing protocols to prepare smooth surfaces with well-defined, regular steps, we now show some preliminary thin film growth results on these unconventional substrates.

\begin{figure}[tb]
    \centering
    \includegraphics[width=\columnwidth]{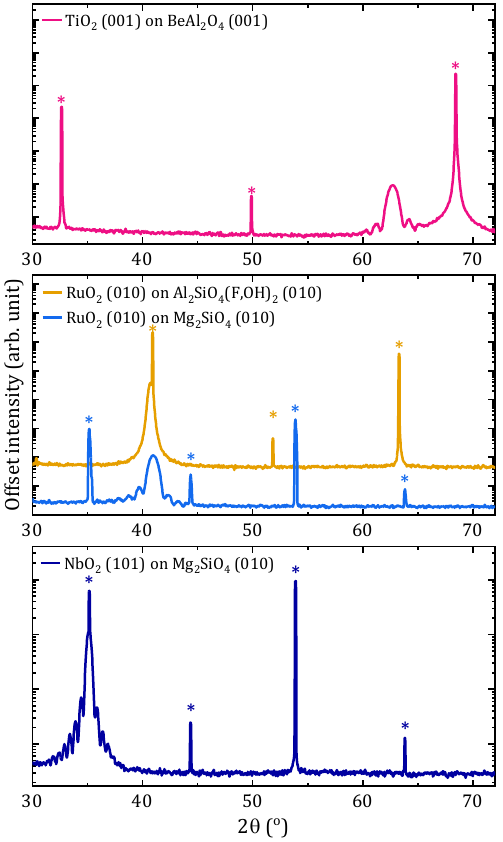}
    \caption{XRD of epitaxial rutile \ce{TiO2}, \ce{RuO2}, and \ce{NbO2} thin films grown on alexandrite, forsterite, and topaz substrates. Substrate peaks are denoted with asterisks.}
    \label{fig:XRD}
\end{figure}

As a first example, we show the epitaxial growth of rutile \ce{RuO2} on \ce{BeAl2O4} (001) and (010).
%As a second example, we demonstrate the growth of \ce{VO2} on topaz (001).
As a second example, we showcase the growth of \ce{NbO2} (101) on \ce{Mg2SiO4} (010).
Symmetrical $\theta-2\theta$ XRD scans of these three epitaxial film examples are shown in Fig. \ref{fig:XRD}. For each film, only peaks belonging to a single family of planes were detected, indicating that the films have a single out-of-plane orientation. 
The in-plane alignment with the substrates, and thus epitaxial growth, was established from the observed RHEED patterns % and $\phi$-scans
(see Fig. \ref{fig:RHEED}).

% One of the frequently encountered challenges in the growth of \ce{VO2} on \ce{TiO2} is interdiffusion across the interface \cite{Muraoka2002, Paik2015}.
% Y. Muraoka and Z. Hiroi, Appl. Phys. Lett. 80, 583 (2002).

The detailed study of the microstructure and the physical properties of the differently strained \ce{RuO2} layers will be the topic of a future article.

%\textcolor{red}{Darrell, on purpose we keep this extremely short with no phi scans or RSMs so that Luka can follow up with a dedicated article on unconventional strain-engineering of \ce{RuO2}. If the reviewers ask for it, we can deliver...}

\section{Conclusion}
In summary, we have presented three promising single-crystalline substrates that can be thermo-chemically prepared to serve as substrates for epitaxial thin film growth.
The ability to produce smooth, regularly-stepped substrates provides an ideal starting ground for the exploration of heteroepitaxial thin film growth. 

As a first case study, we briefly illustrated the single-oriented epitaxial growth of \ce{NbO2} (101) on \ce{Mg2SiO4} (010) and \ce{RuO2} (001) on \ce{BeAl2O4} (001) as well as topaz (001), (010) and \ce{Mg2SiO4} (010), opening an unexplored avenue for heteroepitaxy of rutile-structured materials on alexandrite-structured substrates.

The best results thus far have been achieved with \ce{Mg2SiO4} and \ce{BeAl2O4}, whereas achieving a smooth termination of topaz remains an open challenge.
In particular, \ce{BeAl2O4} stands out due to its exceptional thermal and chemical stability, increasing its versatility as a substrate for rutiles with a wide variety of optimal growth conditions. 

%Future work could investigate solid solutions to tune the lattice parameters of alexandrite , e.g., substituting Al in \ce{BeAl2O4} with Cr \cite{Tabata1981} or substituting Mg in \ce{Mg2SiO4} with Fe or Mn.\cite{Finch1980, Kanazawa2007 }

While the case studies here examined the growth of rutile, the substrates presented here are also very well suited for the growth of other crystalline materials, in particular materials with the olivine structure.

%Ultimately, the most versatile substrate for rutile thin films would be a new rutile substrate. It is worth revisiting bi-rutile and tri-rutile compounds with the formulae \ce{ABO4} and \ce{AB2O6}, respectively. 

\section{\label{sec:methods}Methods}
Atomic force microscopy images were recorded \textit{ex situ} in air at room temperature using an Asylum Research Cypher S instrument with NanoWorld Arrow-UHFAuD-10 cantilever probes. Post-processing was performed using Gwyddion (v.2.67)

\textit{In situ} laser annealing was performed inside a Veeco Gen10 molecular-beam epitaxy chamber equipped with a mid-infrared (\SI{10.6}{\micro\meter}) \ce{CO2} laser substrate-heating system built by Epiray. The temperature was controlled using a \SI{7.5}{\micro\meter} wavelength pyrometer probing the backside of the substrate, which was calibrated to the melting point of sapphire. A ramp rate of \SI{200}{\celsius\per\minute} was used up to \SI{1000}{\celsius} and \SI{100}{\celsius\per\minute} thereafter. This \textit{in situ} annealing protocol follows the example of \citet{Braun2020}.
Distilled ozone was supplied about 41 mm from the center of the front side of the substrate via a water-cooled stainless steel nozzle and the chamber background pressure was kept at \SI{1E-6}{\Torr}. The base pressure of the chamber is better than \SI{1E-8}{\Torr}.

Rutile \ce{RuO2} thin films were grown by electron-beam evaporation of ruthenium in an ambient of distilled ozone. \ce{NbO2} thin films were grown using suboxide molecular-beam epitaxy using a \ce{Nb2O5} charge (H.C. Stark \SI{99.99}{\percent} contained in an iridium crucible in a high-temperature MBE effusion cell, as has been used to grow \ce{KNbO3} by MBE.\cite{Hazra2024} 
% S. Hazra, T. Schwaigert, A. Ross, H. Lu, U. Saha, V. Trinquet, B. Akkopru-Akgun, B.Z. Gregory, A. Mangu, S. Sarker, T. Kuznetsova, S. Sarker, X. Li, M.R. Barone, X. Xu, J.W. Freeland, R. Engel-Herbert, A.M. Lindenberg, A. Singer, S. Trolier-McKinstry, D.A. Muller, G. Rignanese, S. Salmani-Rezaie, V.A. Stoica, A. Gruverman, L. Chen, D.G. Schlom, and V. Gopalan, “Colossal Strain Tuning of Ferroelectric Transitions in KNbO3 Thin Films,” Adv. Mater. 36(52), 2408664 (2024).
The dominant vapor pressure emanating from a \ce{Nb2O5} charge is tetravalent \ce{NbO2}.\cite{Adkison2020} This suboxide approach avoids the need (and associated flux instability) for electron-beam evaporation of Nb metal. Suboxide MBE was performed at base pressure without added oxidant.
\textit{In situ} RHEED images were collected using a Staib electron gun and kSA camera system at a beam energy of \SI{13}{\kilo\electronvolt}.

Furnace annealing was performed in an alumina tube with a flow of \SI{60}{\milli\liter\per\hour} oxygen gas at atmospheric pressure.
Prior to annealing, all substrates were cleaned using isopropanol and dried with nitrogen.
Additionally, the topaz samples were etched in a buffered HF solution for \SI{60}{seconds}.

X-ray diffraction was performed using a PANalytical Empyrean diffractometer, equipped with a hybrid mirror-monochromator and a PIXcel detector.

\begin{acknowledgments}

The authors thank Maya Ramesh, Anna S. Park, Tomas A. Kraay, and Steven Button for experimental assistance.
% The authors thank Oxide Corp., Japan, Surfacenet, Germany, CrysTec, Germany, and Synoptics, USA 
%The authors acknowledge the use of facilities and instrumentation supported by the National Science Foundation (NSF) through the Cornell University Materials Research Science and Engineering Center DMR-1719875. The Helios FIB is supported by NSF DMR-1539918.
This work made use of the thin film facility of the Platform for the Accelerated Realization, Analysis, and Discovery of Interface Materials (PARADIM), which is supported by the NSF under Cooperative Agreement No. DMR-2039380.
M.K. acknowledges support through the REU Site: Summer Research Program at PARADIM, which is supported by the NSF under Cooperative Agreement No. DMR-2150446.
Y.A.B. acknowledges support from the Dutch Research Council (NWO) through the project \textit{Conductivity on demand -- turning an insulator into a metal} with project number 019.223EN.017 of the research program RUBICON.
%Y.A.B. also acknowledges support from by SUPREME, one of seven centers in JUMP 2.0, a Semiconductor Research Corporation (SRC) program sponsored by DARPA.
L.B.M. was supported as part of the Center for Electrochemical Dynamics and Reactions at Surfaces (CEDARS), an Energy Frontier Research Center funded by the U.S. Department of Energy, Office of Science, Office of Basic Energy Sciences, under Award DE-SC0023415.

% Moore foundation?

\end{acknowledgments}

%%%%%%%%%%%%    AUTHOR CONTRIBUTIONS
\section{Author declarations}
\subsection{Conflict of Interest}
The author D.G.S. has been granted U.S. Patent No. 11,462,402
(4 October 2022) with the title “Suboxide Molecular-Beam Epitaxy
and Related Structures.”

\subsection{Author contributions}
M.K. and Y.A.B. contributed equally.
M.K. performed and analyzed the AFM under guidance of Y.A.B.
Y.A.B. conceived the project and fabricated the samples together with L.B.M under the guidance of D.G.S., who suggested these rutile-related substrates.\
%N.S. performed the STEM analysis with guidance from D.A.M.
M.K and Y.A.B. wrote the manuscript with input from all authors.

% All authors read and approved of the manuscript

%%%%%%%%%%%%%

\section*{Data Availability Statement}

The data that support the findings of this study are available within the article. Additional data related to film growth and structural characterization by RHEED, XRD and AFM are available at 
\url{https://doi.org/10.34863/qzna-rk32.} %% CHANGE TO UNIQUE ONE
Any additional data connected to the study are available from the corresponding author upon reasonable request.

%\nocite{*} % prints ALL entries of bib file
\bibliography{bib20251106_v2}% Produces the bibliography via BibTeX.

~
\clearpage

\onecolumngrid

\appendix

\section{Appendixes}

\begin{figure}[h]
    \centering
    \includegraphics[width=0.7\linewidth]{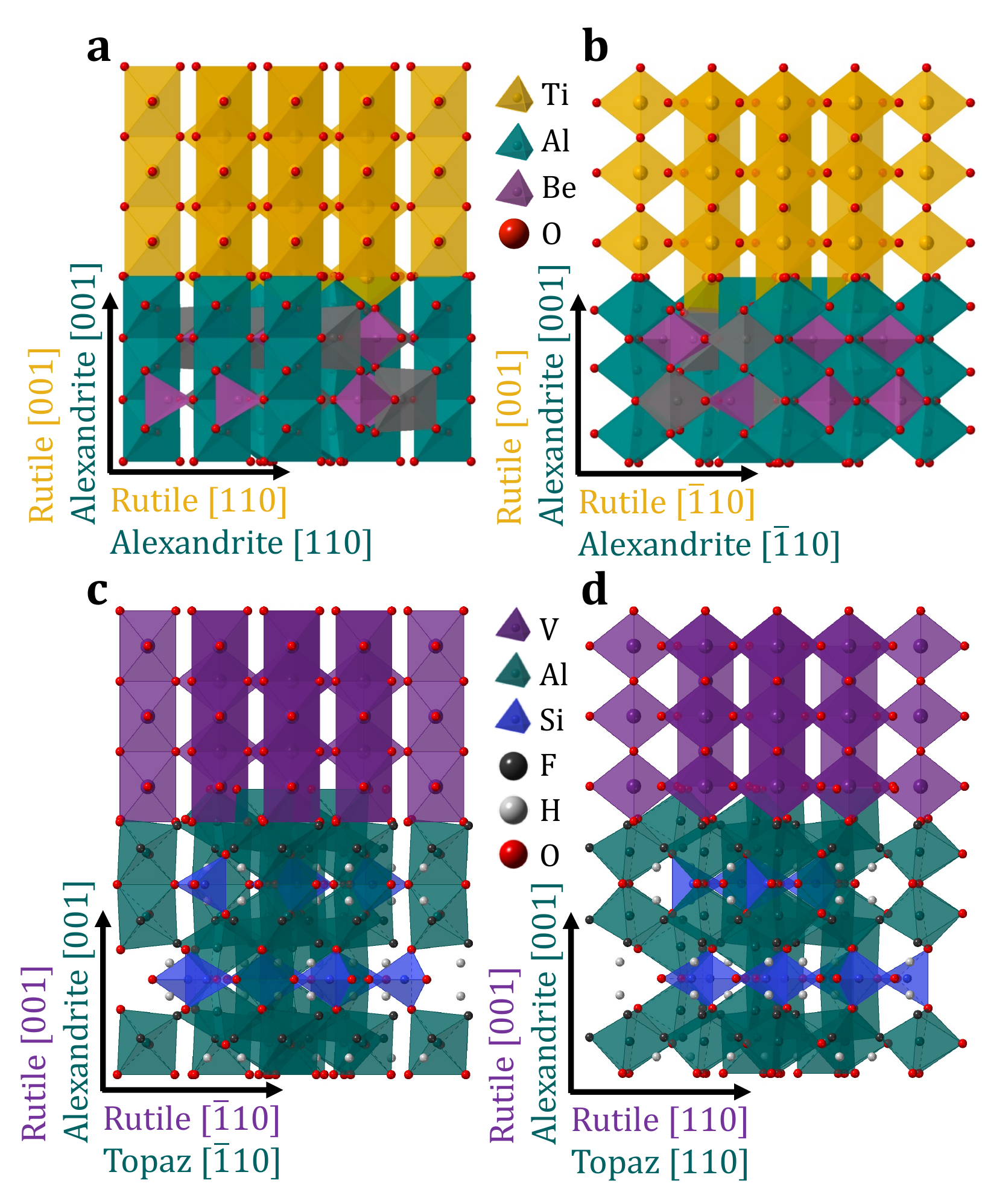}
    \caption{Cross-sectional schematic of (a,b) \ce{TiO2} (001) on \ce{BeAl2O4} (001) and (c,d) \ce{VO2} on topaz (001). emphasizing the continuation of the edge-sharing octahedral network along the $c$-axis. Complementary views to Fig. \ref{fig:RuO2_001_cartoon} rotated \SI{45}{\degree} around the $c$-axis.}
    \label{fig:SI_001_along110}
\end{figure}

\begin{figure} [h]
    \centering
    \includegraphics[width=\columnwidth]{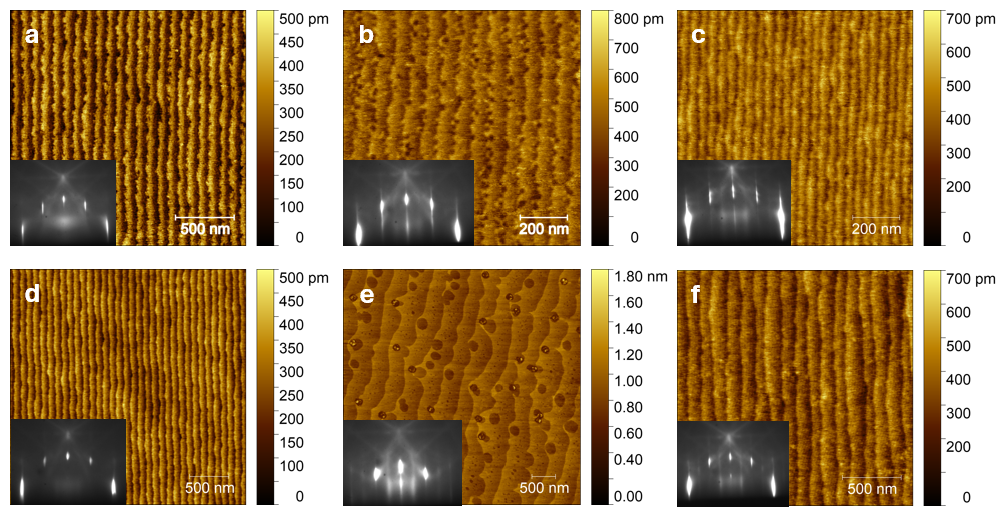}
    \caption{AFM scans and RHEED patterns of (001) \ce{BeAl2O4}  (a) in the as-received state and after (b) laser annealing at \SI{1100}{\celsius} for \SI{200}{\second}, (c) \SI{1200}{\celsius} for \SI{200}{\second}, (d) \SI{1250}{\celsius} for \SI{200}{\second}, (e) \SI{1300}{\celsius} for \SI{200}{\second}, and (f) \SI{1400}{\celsius} for \SI{200}{\second}.}
    \label{fig:001BAO_supp}
\end{figure}

\begin{figure} [h]
    \centering
    \includegraphics[width = 0.45\columnwidth]{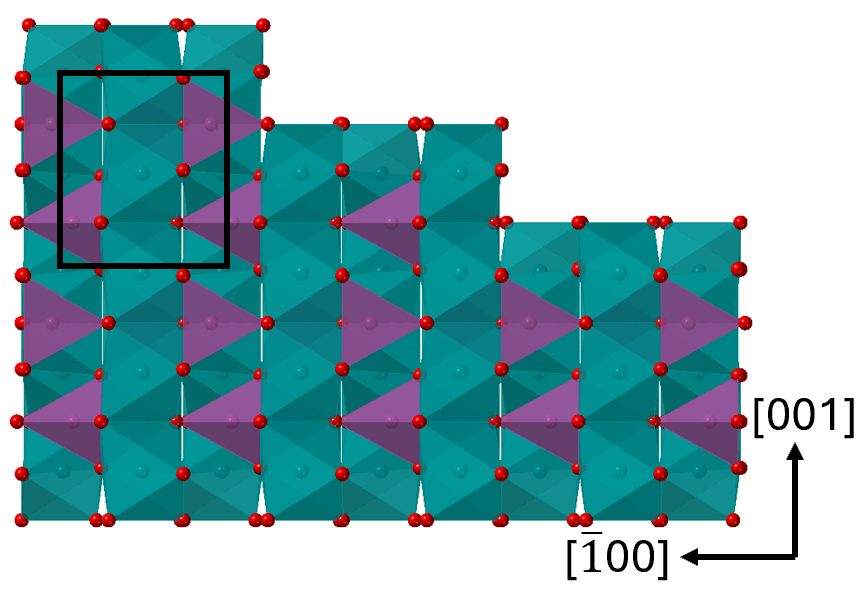}
    \caption{Cross-sectional schematic of the possible surface termination of a vicinal (001) \ce{BeAl2O4} substrate with 1/2 unit cell step height.}
    \label{fig:001BAO_steps}
\end{figure}

\begin{figure} [h]
    \centering
    \includegraphics[width=\columnwidth]{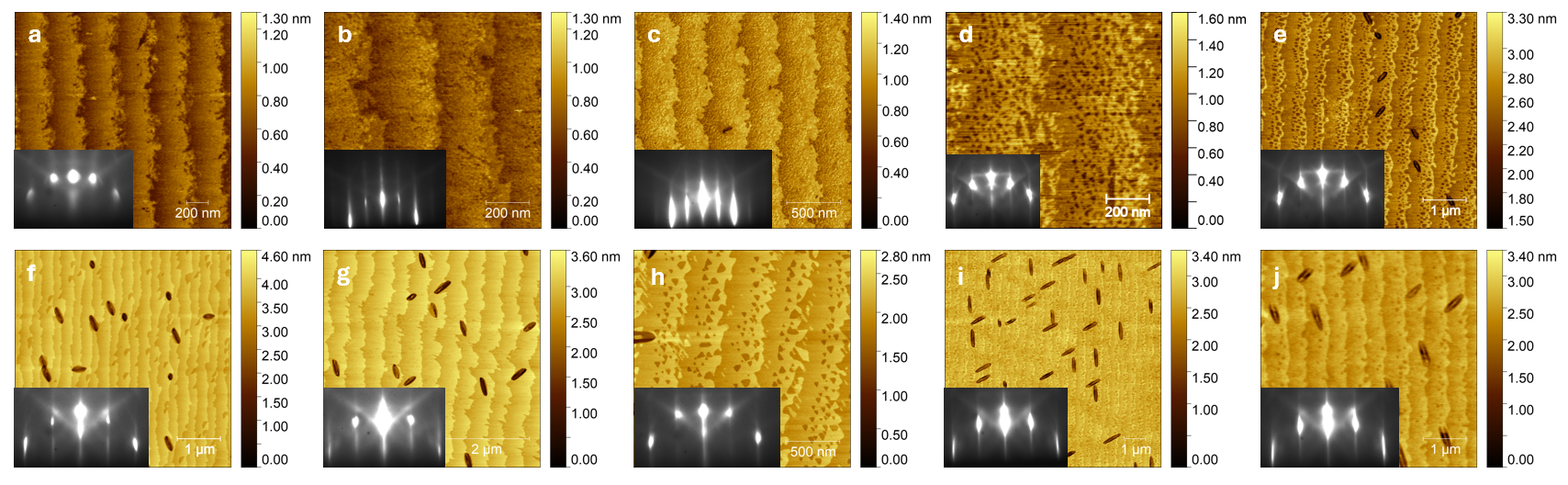}
    \caption{AFM scans and RHEED patterns of (010) \ce{BeAl2O4}  (a) in the as-received state and after (b) laser annealing at \SI{1100}{\celsius} for \SI{200}{\second}, (c) \SI{1200}{\celsius} for \SI{200}{\second}, (d) \SI{1300}{\celsius} for \SI{200}{\second}, (e) \SI{1350}{\celsius} for \SI{200}{\second}, (f) \SI{1350}{\celsius} for \SI{320}{\second}, (g) \SI{1400}{\celsius} for \SI{200}{\second}, (h) \SI{1450}{\celsius} for \SI{200}{\second}, (i) \SI{1450}{\celsius} for \SI{500}{\second}, and (j) \SI{1500}{\celsius} for \SI{200}{\second}.}
    \label{fig:010BAO_supp}
\end{figure}

\begin{figure} [h]
    \centering
    \includegraphics[width=\columnwidth]{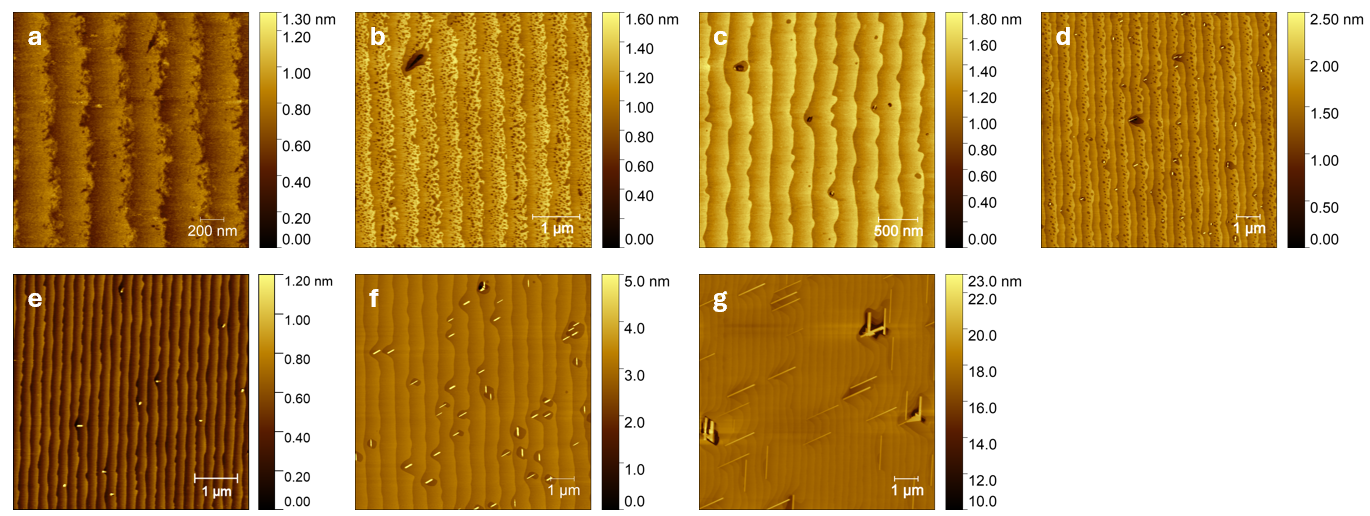}
    \caption{AFM scans of (010) \ce{BeAl2O4}  (a) in the as-received state and after (b) furnace annealing at \SI{850}{\celsius} for \SI{2}{\hour}, (c) \SI{900}{\celsius} for \SI{1}{\hour}, (d) \SI{950}{\celsius} for \SI{1}{\hour}, (e) \SI{1000}{\celsius} for \SI{1}{\hour}, (f) \SI{1100}{\celsius} for \SI{1}{\hour}, and (g) \SI{1200}{\celsius} for \SI{1}{\hour}.}
    \label{fig:010BAO_T_supp}
\end{figure}

\begin{figure} [h]
    \centering
    \includegraphics[width = 0.45\columnwidth]{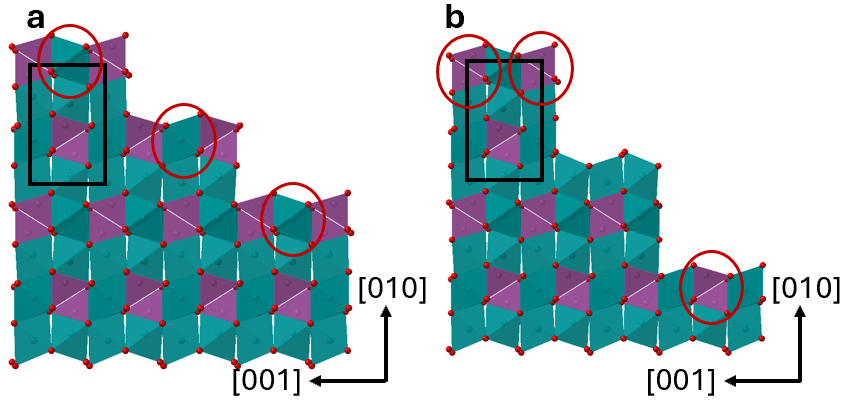}
    \caption{Cross-sectional schematic of possible surface terminations of vicinal (010) \ce{BeAl2O4} substrates with (a) 2/5 unit cell step height and (b) 3/5 unit cell step height. In (a), the aluminum octahedra face in two different directions on alternating steps, which creates the double termination. In (b), the beryllium dimers only appear in every other step surface, also creating a double termination.}
    \label{fig:010BAO_steps}
\end{figure}

\begin{figure} [h]
    \centering
    \includegraphics[width=\columnwidth]{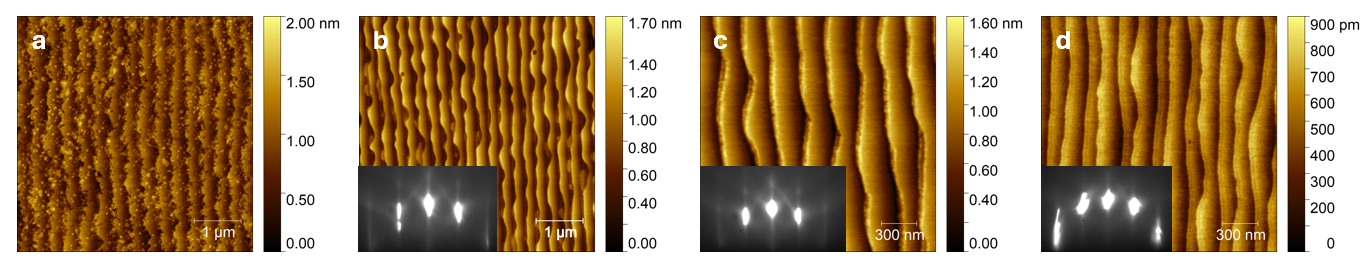}
    \caption{AFM scans and RHEED patterns of (010) \ce{Mg2SiO4}  (a) in the as-received state and after (b) laser annealing at \SI{1200}{\celsius} for \SI{200}{\second}, (c) \SI{1300}{\celsius} for \SI{200}{\second}, and (d) \SI{1400}{\celsius} for \SI{200}{\second}.}
    \label{fig:010MSO_supp}
\end{figure}

\begin{figure} [h]
    \centering
    \includegraphics[width=\columnwidth]{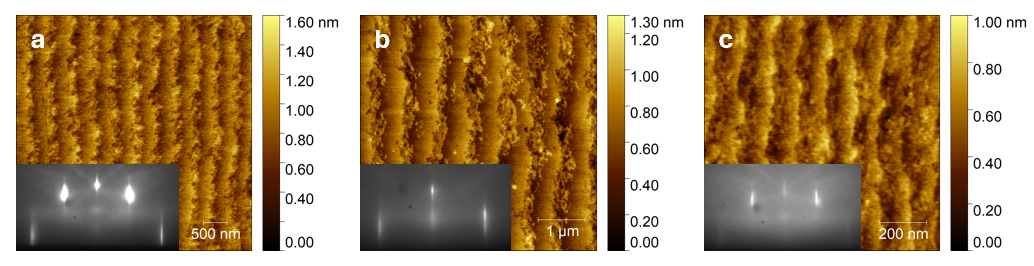}
    \caption{AFM scans and RHEED patterns of (001) \ce{Al2SiO4(\text{F,OH})_2}  (a) after BHF etching followed by (b) laser annealing at \SI{700}{\celsius} for \SI{200}{\second}, and (c) \SI{800}{\celsius} for \SI{200}{\second}.}
    \label{fig:001Topaz_supp}
\end{figure}

\begin{figure} [h]
    \centering
    \includegraphics[width=\columnwidth]{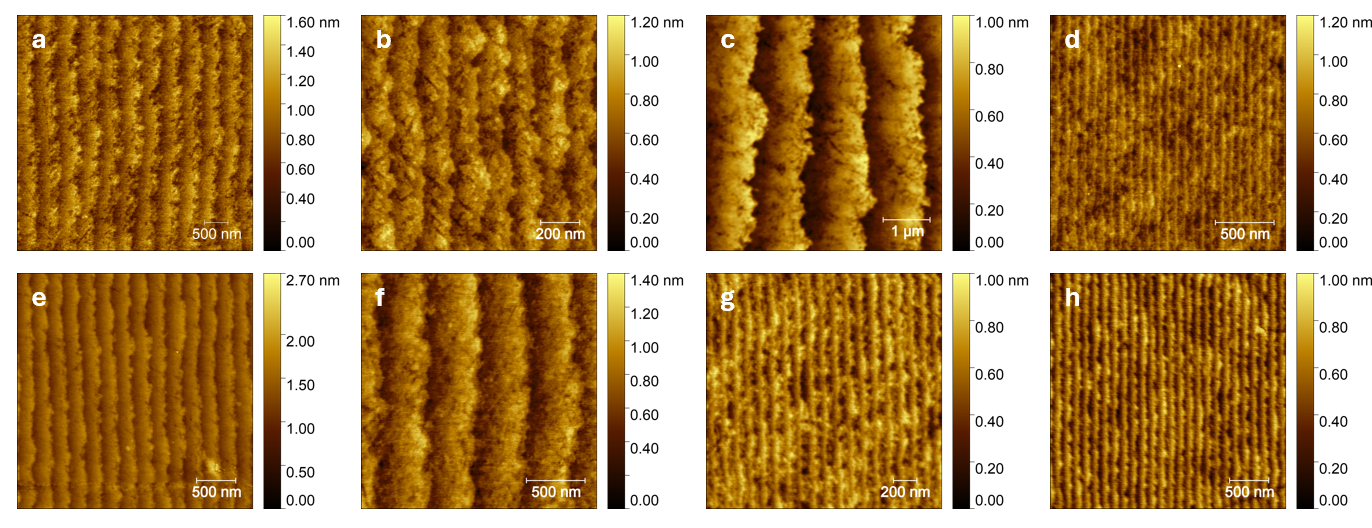}
    \caption{AFM scans of (001) \ce{Al2SiO4(\text{F,OH})_2}  (a) in the as-received state and after (b) furnace annealing at \SI{550}{\celsius} for \SI{2}{\hour}, (c) \SI{600}{\celsius} for \SI{1}{\hour}, (d) \SI{650}{\celsius} for \SI{1}{\hour}, (e) \SI{700}{\celsius} for \SI{1}{\hour}, (f) \SI{750}{\celsius} for \SI{1}{\hour}, (g) \SI{800}{\celsius} for \SI{1}{\hour}, and (h) \SI{850}{\celsius} for \SI{1}{\hour}.}
    \label{fig:001Topaz_T_supp}
\end{figure}

\begin{figure} [h]
    \centering
    \includegraphics[width=\columnwidth]{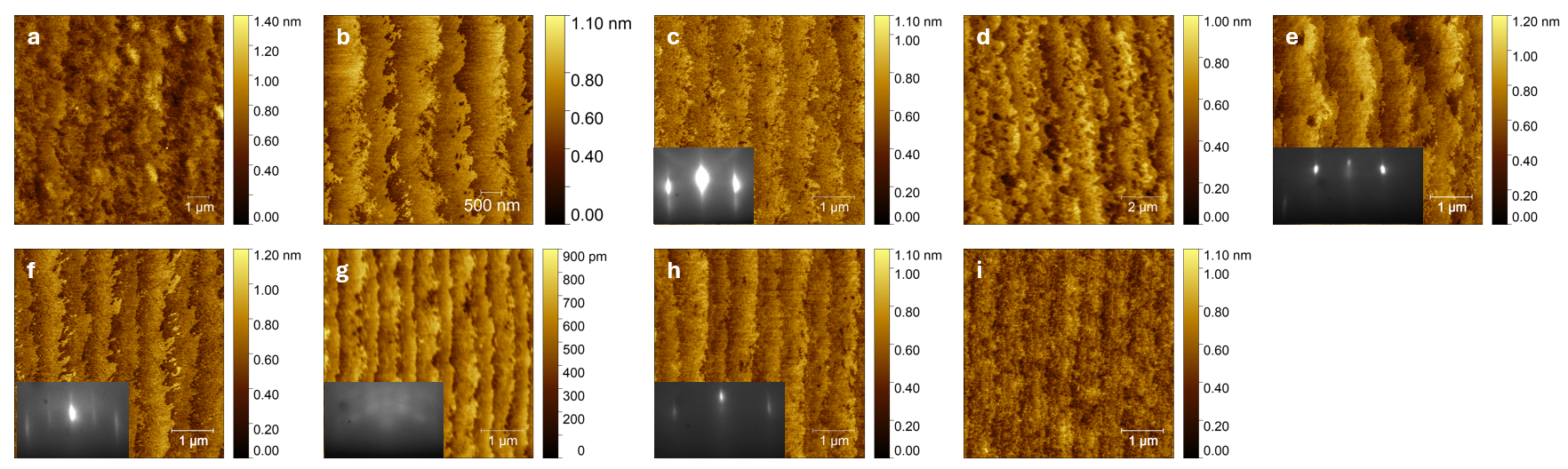}
    \caption{AFM scans and RHEED patterns of (010) \ce{Al2SiO4(\text{F,OH})_2}  (a) in the as-received state and after (b) etching in BHF followed by (c) laser annealing at \SI{700}{\celsius} for \SI{200}{\second}, (d) \SI{750}{\celsius} for \SI{500}{\second}, (e) \SI{800}{\celsius} for \SI{200}{\second}, (f) \SI{850}{\celsius} for \SI{200}{\second}, (g) \SI{850}{\celsius} for \SI{500}{\second}, (h) \SI{900}{\celsius} for \SI{200}{\second}, and (i) \SI{1000}{\celsius} for \SI{200}{\second}.}
    \label{fig:010Topaz_supp}
\end{figure}

\begin{figure} [h]
    \centering
    \includegraphics[width=\columnwidth]{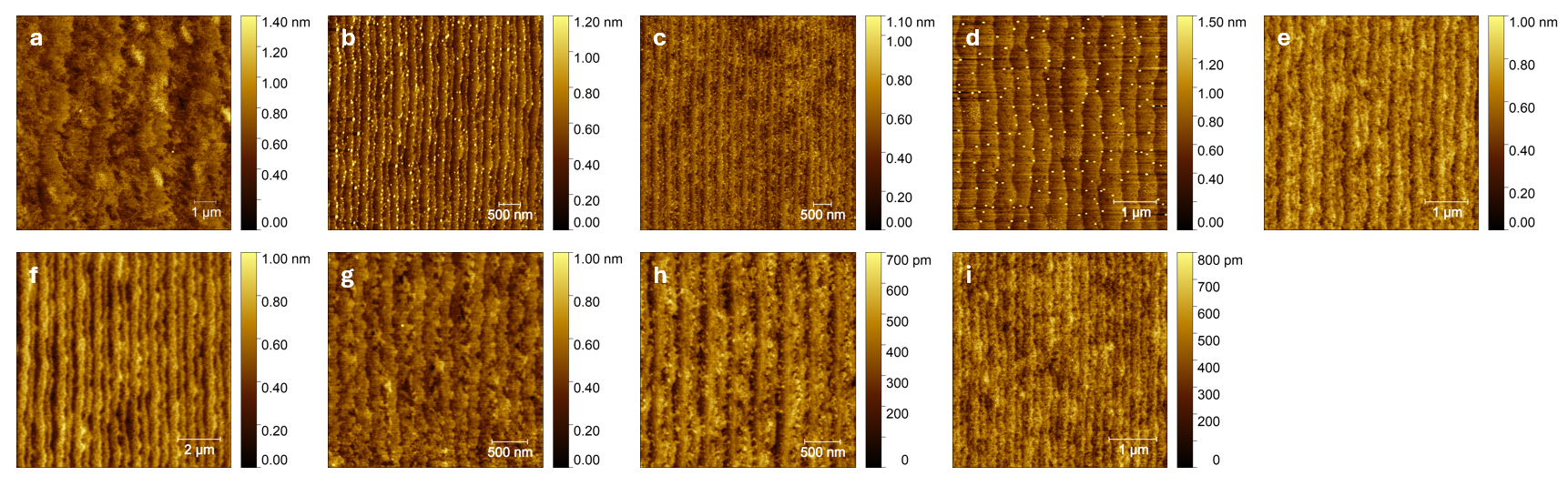}
    \caption{AFM scans of (010) \ce{Al2SiO4(\text{F,OH})_2}  (a) in the as-received state and after (b) furnace annealing at \SI{550}{\celsius} for \SI{2}{\hour}, (c) \SI{575}{\celsius} for \SI{1}{\hour}, (d) \SI{600}{\celsius} for \SI{1}{\hour}, (e) \SI{650}{\celsius} for \SI{1}{\hour}, (f) \SI{700}{\celsius} for \SI{1}{\hour}, (g) \SI{750}{\celsius} for \SI{1}{\hour}, (h) \SI{800}{\celsius} for \SI{1}{\hour}, and (i) \SI{850}{\celsius} for \SI{1}{\hour}.}
    \label{fig:010Topaz_T_supp}
\end{figure}

\begin{figure} [h]
    \centering
    \includegraphics[width=0.7\columnwidth]{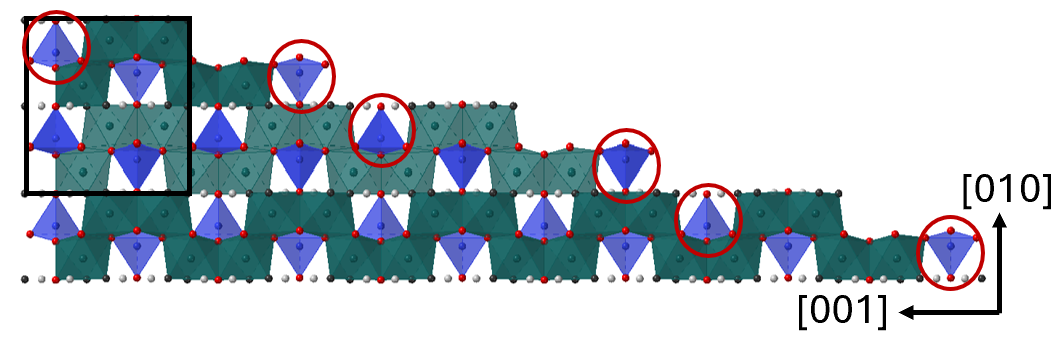}
    \caption{Cross-sectional schematic of the possible surface termination of a vicinal (010) topaz substrate with quarter-unit-cell steps. The silicon coordination polyhedra face (circled) opposite directions on alternating steps, creating a quadruple termination.}
    \label{fig:010Topaz_steps}
\end{figure}
%% are the octahedrons also rotated?

%% Is it okay to talk about dimers for these units?

\begin{figure}[h]
    \centering
    \includegraphics[width=0.5\linewidth]{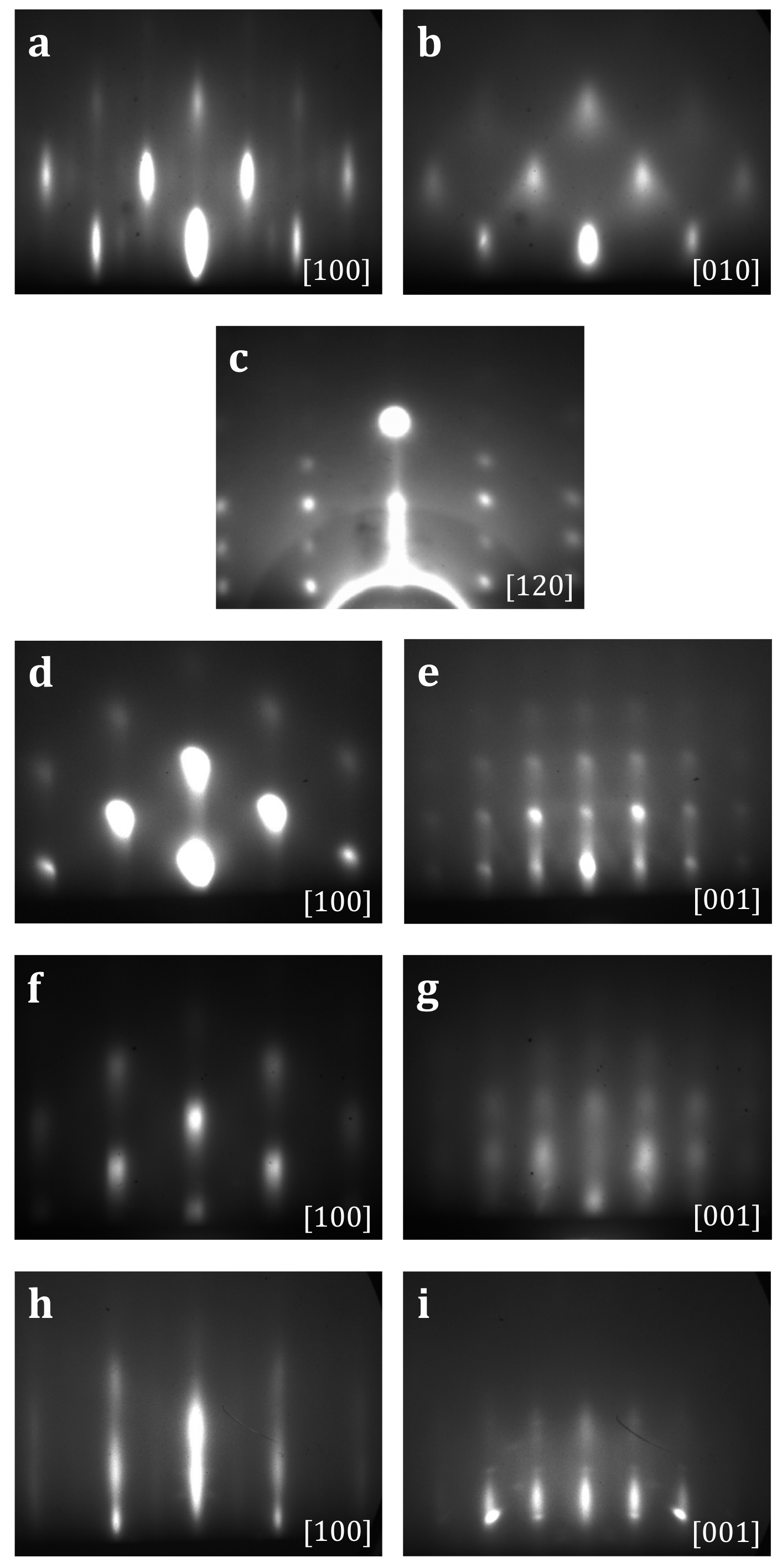}
    \caption{RHEED patterns recorded at the end of the growth of (a,b) \ce{TiO2} (001) on alexandrite (001),  (c) \ce{VO2} (001) on topaz (001), (d,e) \ce{RuO2} (010) on topaz (010), (f,g) \ce{RuO2} (010) on forsterite (010), (h,i) \ce{NbO2} (101) on \ce{Mg2SiO4} (010). The electron beam was aligned with respect to the substrate Miller indices indicated at the bottom right corner of the panels.}
    \label{fig:RHEED}
\end{figure}

\end{document}